%% file: bare_jrnl_new_sample4.tex
\documentclass[letterpaper,journal]{IEEEtran}
\usepackage{amsmath,amsfonts,amssymb}
\usepackage{tabularx}
\usepackage{array}
\usepackage[caption=false,font=normalsize,labelfont=rm,textfont=rm]{subfig}
\usepackage{textcomp}
\usepackage{stfloats}
\usepackage{url}
\usepackage[table]{xcolor}
\usepackage{verbatim}
\usepackage{graphicx}
\usepackage{multirow}
\usepackage[acronym]{glossaries}
\usepackage[hidelinks]{hyperref}
\usepackage[linesnumbered,ruled,vlined]{algorithm2e}
\usepackage{cite}
\loadglsentries{glossaries.tex}

\DeclareMathOperator*{\argmin}{arg\,min}

\begin{document}
	
	\title{LocAttMamba: A Low-Complexity Mamba Framework with Attention-Based Multi-AP Fusion for Indoor Localization}
	
	\author{MohammadMahdi Ghadaksaz, Mohammad Farzanullah,~\IEEEmembership{Graduate Student Member,~IEEE} Melike Erol-Kantarci, ~\IEEEmembership{Fellow,~IEEE}
		\thanks{MohammadMahdi Ghadaksaz, Mohammad Farzanullah, and Melike Erol-Kantarci are with the School of Electrical Engineering and Computer Science, University of Ottawa, Ottawa, ON K1N 6N5, Canada (e-mail: mghad017@uottawa.ca; mfarz086@uottawa.ca; melike.erolkantarci@uottawa.ca).}
	}
	

	
	\maketitle
	\glsresetall[acronym]
	\begin{abstract}
		Accurate and low-complexity indoor localization is important for location-based
		services in \gls{5G} and \gls{6G} networks, where
		positioning devices operate under limited computational budgets and \gls{NLoS} conditions. Indoor localization has been studied widely using traditional signal-level localization approaches. However, these techniques often show degraded performance in \gls{NLoS} scenarios. Recently, \gls{AI}-based techniques, including transformers, have been applied to address
        these challenges. While transformer-based architectures can
		capture the dependencies within the measurements collected from distributed \glspl{AP}, their high computational complexity results in a large number of multiply-accumulate operations and long inference time. In contrast, lightweight recurrent and convolutional models trade this cost for degraded accuracy. In this paper, we propose LocAttMamba, a low-complexity localization framework in which the \gls{CIR} and time-based features of each \gls{AP} are processed by a separate Mamba encoder with near-linear complexity, and the resulting per-\gls{AP} embeddings are fused through a multi-head attention layer that weights each \gls{AP} according to its importance at every time step. The framework jointly predicts the user location and its per-axis uncertainty, which is refined through a post-hoc calibration step. We evaluate the proposed framework using two real-world \gls{5G} and \gls{UWB} measurement datasets. Our numerical results reveal that LocAttMamba obtains a mean \gls{2-D} positioning error of $1.004$ m and $0.599$ m, respectively, on \gls{5G} and \gls{UWB} datasets, outperforming the second-best benchmark by $9.79\%$ and $5.82\%$,  while requiring the fewest multiply-accumulate operations among all evaluated models and being $4.4$--$16\times$ faster than the transformer-based benchmarks.
	\end{abstract}
	
	
	\glsresetall[acronym]
	
	\section{Introduction}
	\IEEEPARstart{T}{he} widespread deployment of \gls{5G} and the ongoing move toward \gls{6G} are pushing wireless communication forward quickly, and they are opening the way for the next generation of location-based use cases, such as \gls{XR}, \gls{V2X} \cite{10136521}, and physical \gls{AI} \cite{11292806}. This calls for localization methods that are accurate, efficient, robust, and low in computational cost, so that the strict \gls{QoS} requirements for the emerging applications can be met. Indoor localization is especially important, since it supports navigation, safety and rescue operations, and personalized services. In addition, most devices have limited computing power and the measurements they collect are affected by noise and changing surrounding conditions.\par 
	In this context, while \gls{GNSS} is infeasible for indoor localization, and conventional, signal-level localization approaches have poor performance in \gls{NLoS}-rich environments, \gls{DL}-based approaches have drawn attention over the past few years \cite{10534778}. In particular, \gls{MLP}-based models can capture complex, non-linear relationships within the input features, \glspl{CNN} using \gls{2-D} or \gls{3-D} inputs can extract temporal and spatial patterns, and \glspl{RNN} and \gls{LSTM} networks can learn the temporal dependencies of a moving user. More recently, transformers \cite{Vaswani2017Attention} have been introduced as a substitute for \glspl{RNN} and \glspl{LSTM} for handling long-range
	dependencies \cite{10.1145/3505244}. However, their quadratic
	complexity in inference can result in high computational costs and memory usage, limiting their practicality in real-time scenarios. \par 
    Beyond accuracy and computational cost, a practical challenge for \gls{DL}-based indoor localization is the degradation of measurement quality. In real deployments, the \gls{CIR} measurements collected by the \glspl{AP} are affected by interference and blockage, and can even be deliberately perturbed at the physical layer \cite{rob_1}. Prior works have shown that corrupted measurements from even a single source can significantly bias the position estimate of a \gls{DL}-based model, and that adaptively weighting the measurement sources helps mitigate the impact of anomalous signals \cite{rob_2}. Moreover, even with clean measurements, the localization error can vary
significantly across locations due to \gls{NLoS} propagation and multipath, while a point estimate alone provides no indication of its reliability. To address this, the authors in \cite{foliadis2022reliable} estimate the uncertainty of each base station and use it to weight their estimates, although the Monte Carlo dropout variant requires multiple forward passes. Closer to our framework, \cite{technologies13090386} predicts an uncertainty alongside each estimate within an attention-based fusion network to reject unreliable ones. However, this uncertainty is not calibrated, which means the predicted uncertainty may not match the actual error.
    \par 
	To overcome the drawbacks of transformers, Mamba \cite{gu2023mamba} has been introduced as an efficient alternative, which has recently been used in several wireless communication-related problems \cite{arafat-IDN, arafat-mamba, Mamba-traffic-mehrabian}. Mamba is based on the \glspl{SSM}, which offer near-linear time complexity. Mamba relies on a selection mechanism in which the state matrices and the time step become functions of the input, allowing the model to keep the relevant parts of the sequence and discard the rest. Furthermore, Mamba considers a hardware-aware selective scan algorithm instead of standard sequential approach, which keeps the training and inference cost low. \par 
	Nonetheless, a single Mamba processes a single sequence along the time dimension, which can limit its ability to capture spatial dependencies from multiple measurement sources. In this case, to capture both temporal and spatial dependency, Mamba can be combined with multi-head attention-based fusion technique \cite{pmlr-v80-ilse18a} to fuse the information from multiple measurement sources. \par
    In this paper, we propose a low-complexity framework in which each AP is processed by a distinct bidirectional Mamba encoder, and the per-AP embeddings are fused through a multi-head attention layer that adaptively down-weights distorted \glspl{AP}. Following \cite{technologies13090386}, we also predict the uncertainty of each estimate, which, unlike \cite{foliadis2022reliable}, \cite{technologies13090386}, is obtained from a single forward pass of a single model and is calibrated through a post-hoc step.
	\subsection{Related Works}
    Indoor localization has been widely studied using conventional, signal-level approaches, including geometric and fingerprinting-based methods \cite{zafari2019survey}. In particular, geometric methods estimate the user location using \gls{ToA}, \gls{TDoA}, \gls{AoA}, or \gls{RSS} measurements, where their performance significantly degrades in \gls{NLoS}-rich environments \cite{guvenc2009survey}. On the other hand, fingerprinting-based methods match the collected measurements with a pre-collected radio map using simple similarity metrics, such as \gls{KNN}, which limits their ability to exploit high-dimensional measurements, e.g., \gls{CIR}, and to generalize beyond the surveyed reference points. \par 
	Due to these limitations, \gls{DL}-based indoor localization approaches have been proposed over the past few years \cite{MLP, MLP_2, bi-lstm, ACCN}. In particular, the authors in \cite{MLP} propose an \gls{MLP}-based indoor localization approach using \gls{MLP} and \gls{1-D} \gls{CNN} and having \gls{RSS} and \gls{CSI} as input features, where they show that their proposed approach obtains sub-meter localization error. Similarly, in \cite{MLP_2}, the authors introduce an \gls{MLP}-Mixer architecture for \gls{CSI}-fingerprinting localization in \gls{mMIMO} \gls{OFDM} systems, where they show the effectiveness and robustness of their proposed algorithm. The authors in \cite{bi-lstm} consider a bidirectional \gls{LSTM} architecture with temporal modeling for indoor localization, using magnetic data collected by a smartphone. The authors show that their proposed approach obtains lower localization error compared to unidirectional \gls{LSTM}, \gls{MLP}, and \gls{GRU} benchmarks. In \cite{ACCN}, the authors investigate an \gls{ACNN} framework for indoor localization using Wi-Fi signals. Here, the authors consider multiple-\glspl{AP} measurement data, where they show the superiority of their design compared to several existing benchmarks, including standard \glspl{CNN}. \par 
	Several recent works explore attention-based \cite{Sec5GLoc, technologies13090386} and transformer-based \cite{L-SwiGLU-T, SwinLoc} indoor localization-based solutions. In \cite{Sec5GLoc}, the authors introduce secure \gls{CNN}-attention-based solution for indoor localization, where they consider \gls{CIR}, \gls{TDoA}, and \glspl{AP}' position as input features. In this work, the authors compare their proposed solution against \gls{KNN}-based and conventional geometric \gls{TDoA} multilateration-based solutions, confirming the superiority of their framework. The authors in \cite{technologies13090386} introduce an uncertainty-aware multi-modal localization framework using vision and LiDAR modalities and \gls{CNN}-attention mechanism, where predicted uncertainties are used to reject unreliable pose estimates, improving accuracy on real-world robot datasets. The authors in \cite{SwinLoc} investigate a heavy-weight transformer-based approach for indoor localization, obtaining lower \gls{RMSE} over several \gls{DL}-based solutions. Likewise, in \cite{L-SwiGLU-T}, the authors explore light to medium weight transformers to reduce the computational complexity of the heavy-weight transformer architecture, while maintaining the indoor localization performance. \par 
    More recently, Mamba has also been explored for indoor localization \cite{MambaLoc, suwannaphong2025tinyml, nguyen2025uloc, yang2025grammamba}. The authors in \cite{MambaLoc} propose a Mamba-based indoor localization framework for \glspl{UV}, where a transformer teacher trained on \gls{UWB} data distills its knowledge into a \gls{CSI}-based Mamba student, yielding lower \gls{MAE} than a plain Mamba while reducing multiply-accumulate operations and inference time with respect to the transformer. In \cite{suwannaphong2025tinyml}, the authors investigate transformer- and Mamba-based models for on-device indoor localization using Bluetooth and Wi-Fi \gls{RSS} measurements, where the localization is formulated as a room-level classification task. The authors in \cite{nguyen2025uloc} propose a Mamba-based framework that learns the ranging patterns of \gls{UWB} measurements in large-scale environments, obtaining higher accuracy than state-of-the-art approaches. Similarly, the authors in \cite{yang2025grammamba} use Mamba for multi-modal sensor fusion, achieving lower error than the baselines on an indoor positioning dataset.
	\subsection{Motivations \& Contributions}
	Most existing research on indoor localization has relied on \gls{MLP}, recurrent, convolutional, and transformer-based architectures, while the potential of Mamba has only recently started to be explored. This observation is evident in several prior works \cite{MLP, MLP_2, bi-lstm, ACCN, technologies13090386, Sec5GLoc, L-SwiGLU-T, SwinLoc}. In particular, studies such as \cite{MLP, MLP_2} rely on simple \gls{MLP}-based solutions, which often fall short in more complex settings with multiple measurement \glspl{AP}. Other works (i.e., \cite{bi-lstm, ACCN}) adopt recurrent and convolutional architectures that obtain low inference time; however, they suffer from poor localization performance while still requiring a large number of multiply-accumulate operations and model parameters. On the other hand, although the authors in \cite{Sec5GLoc, L-SwiGLU-T, SwinLoc} employ attention- and transformer-based architectures to improve the accuracy, these solutions come with high computational complexity and multiply-accumulate operations, resulting in longer inference time, which complicates their deployment on positioning devices with limited computational power. Furthermore, most of the aforementioned works provide only a point estimate of the user location without quantifying its confidence, whereas the works \cite{foliadis2022reliable} and \cite{technologies13090386}, which consider uncertainty, do not calibrate it. Lastly, although Mamba has recently been adopted for indoor localization, the
    existing works come with several limitations. In particular, \cite{MambaLoc} considers a cross-modal teacher model during training, which requires synchronized measurements from an additional modality, while \cite{suwannaphong2025tinyml} formulates the localization as a room-level classification task rather than estimating the user coordinates. Similarly, \cite{nguyen2025uloc} uses \gls{UWB} range measurements and does not exploit the multipath information in the \gls{CIR}, whereas \cite{yang2025grammamba} focuses on aligning heterogeneous sensing modalities instead of multiple sources of the same modality, such as distributed \glspl{AP}. More importantly, none of these works provides a mechanism to adaptively weight the contribution of individual \glspl{AP}, which is essential when the measurements of some \glspl{AP} are degraded. To the best of our knowledge, this is the first time that a Mamba-based encoder combined with attention-based fusion has been explored for indoor localization using real-world \gls{5G} and \gls{UWB} measurements.
	
	The main contributions of this work are as follows:
	\begin{enumerate}
		\item We propose LocAttMamba, a novel low-complexity indoor localization framework in which the measurements from each \gls{AP} are processed by a dedicated encoder built from stacked forward and reverse Mamba blocks, preserving the near-linear complexity of \glspl{SSM} and thereby keeping the multiply-accumulate operations and the inference time low. The resulting per-\gls{AP} embeddings are subsequently fused through a multi-head attention layer using a shared learnable query, which assigns a weight to each \gls{AP} at every time step according to its importance, rather than treating all \glspl{AP} equally. This is the first work to combine Mamba-based encoders with attention-based \gls{AP} fusion for indoor localization.
		
		\item We evaluate the robustness of the proposed framework under degraded \gls{CIR} quality, considering both single-\gls{AP} and multiple-\gls{AP} distortions over a wide range of \gls{SNR} values. In addition, we analyze the learned attention weights and show that the fusion layer adaptively down-weights the distorted \glspl{AP}, which provides a degree of interpretability to the fusion mechanism and explains the graceful degradation observed under noisy conditions.
		
		\item We conduct an extensive evaluation on two real-world datasets, namely a \gls{5G} and a \gls{UWB} indoor localization dataset, against several benchmarks spanning conventional \gls{ML}, recurrent, convolutional, transformer-, and Mamba-based approaches. Beyond the localization error, we also report the multiply-accumulate operations and the inference time of every benchmark, together with an ablation study over the fusion strategy, the input features, the sequence model, and the training objective, in order to isolate the contribution of each design choice.
	\end{enumerate}
	
	Our numerical results validate the effectiveness of the proposed framework. In particular, LocAttMamba achieves a mean error of $1.004$ m on the \gls{5G} dataset and $0.599$ m on the \gls{UWB} dataset, outperforming the second-best benchmark by $9.79\%$ and $5.82\%$, respectively, while requiring the fewest multiply-accumulate operations among all evaluated models, i.e., $0.446$ M/sample and $0.953$ M/sample, and achieving the third-lowest inference time, which is $4.4$--$16\times$ faster than the transformer-based benchmarks. Moreover, the post-hoc calibration reduces the normalized squared error statistic from $34.6$ to $2.15$, close to the ideal value of $2$ for a two-dimensional Gaussian. Finally, under a fixed \gls{SNR} of $5$ dB with three distorted \glspl{AP} on the \gls{5G} dataset, the proposed framework obtains $37.85\%$, $74.80\%$, $67.76\%$, and $9.64\%$ lower mean error compared to \cite{MambaLoc}, \cite{Sec5GLoc}, \cite{L-SwiGLU-T}, and \cite{bi-lstm}, respectively, confirming that the attention-based fusion effectively mitigates the impact of unreliable measurement sources.
	\subsection{Organization}
	The remainder of this paper is organized as follows: Section \ref{sec:problem} presents the discussed problem followed by the methodology in Section \ref{sec:methodology}.
	Section \ref{sec:NUM} provides the numerical results and analysis, and finally Section \ref{sec:Conclusions} concludes the paper with conclusions and future works.  \par 

	\section{PROBLEM DESCRIPTION}
	\label{sec:problem}
	In this section, we formulate the user localization problem using \gls{CIR} and time-related feature measurements from multiple \glspl{AP}. Let us assume a dataset contains $T$ time steps of \gls{CIR} measurement with $N_{\text{\gls{AP}}}$ \glspl{AP}. We denote the complex \gls{CIR} at time step $t$ with $\nu$ elements  measured by \gls{AP} $i$ as:
	\begin{equation}
		\mathbf{h}_{i}^{(t)} = \left[h_{i, 1}^{(t)}, h_{i, 2}^{(t)}, \cdots, h_{i, \nu}^{(t)}\right]^\top \in \mathbb{C}^{\nu},
	\end{equation}
	where $h_{i,k}^{(t)} \in \mathbb{C}$ is the $k$-th element of the measured \gls{CIR} by \gls{AP} $i$ at time step $t$, for all $k \in \left\{1, 2, \cdots, \nu\right\}$. Furthermore, we assume the availability of the estimated \gls{TDoA} and \gls{ToA}-offset, denoted as $\xi_i^{\left(t\right)} \in \mathbb{R}$ and  $\zeta_i^{\left(t\right)}  \in \mathbb{R}$, respectively. A set of $W$ time steps of measurement starting at time step $t$ for \gls{AP} $i$, denoted by $\mathbf{O}_i^{(t)}$, is then given by:
	\begin{equation}
		\begin{split}
			\mathbf{O}_i^{(t)} = \Big\{&
			\left[\mathbf{h}_i^{(t)}, \xi_i^{(t)}, \zeta_i^{(t)}\right],
			\left[\mathbf{h}_i^{(t+1)}, \xi_i^{(t+1)}, \zeta_i^{(t+1)}\right], \\
			&
			\ldots,
			\left[\mathbf{h}_i^{(t+W-1)}, \xi_i^{(t+W-1)}, \zeta_i^{(t+W-1)}\right]
			\Big\}.
		\end{split}
	\end{equation}
	To obtain a real-valued representation, we concatenate the real and
	imaginary parts of the \gls{CIR} with the two time-related features into
	a single feature vector $\mathbf{z}_i^{(t)} \in \mathbb{R}^{2\nu+2}$ for each time step: 
	\begin{equation}
		\mathbf{z}_i^{(t)} = \left[\Re(\mathbf{h}_i^{(t)})^\top,\ \Im(\mathbf{h}_i^{(t)})^\top,\ \xi_i^{(t)},\ \zeta_i^{(t)}\right]^\top \in \mathbb{R}^{2\nu+2},
	\end{equation}
	where $\Re\left(.\right)$ and $\Im\left(.\right)$ are the real and imaginary operators, respectively. Subsequently, the observation matrix $\mathbf{Z}_{i}^{(t)}$, containing the $W$ time steps of $\mathbf{z}_i$ vectors, starting from time step $t$ is expressed as:
	\begin{equation}
		\label{eq:observation_matrix}
		\mathbf{Z}_i^{(t)} = \left[\mathbf{z}_i^{(t)}, \mathbf{z}_i^{(t+1)}, \cdots, \mathbf{z}^{(t+W-1)}_i\right]^\top \in \mathbb{R}^{W \times (2\nu+2)}.
	\end{equation}
	\par 
	In a similar manner, the \gls{2-D} location of the user during the $W$ time steps window starting at time step $t$ can be written as:
	\begin{equation}
		\mathbf{L}^{\left(t\right)} = 
		\begin{bmatrix}
			l_x^{(t)} & l_y^{(t)} \\ 
			l_x^{(t+1)} & l_y^{(t+1)} \\
			\vdots & \vdots \\
			l_x^{(t+W-1)} & l_y^{(t+W-1)}
		\end{bmatrix}
		\in \mathbb{R}^{W \times 2},
	\end{equation}
	where $l_x^{(t)}$ and $l_y^{(t)}$ represent the user location at time step $t$ over x-axis and y-axis, respectively. \par
	Equation (\ref{eq:observation_matrix}) can be applied to all \glspl{AP} for $i \in \left\{1, 2, \cdots, N_{\text{AP}}\right\}$ to construct synchronized matrices $\mathbf{Z}_1^{(t)}, \mathbf{Z}_2^{(t)}, \cdots, \mathbf{Z}_{N_{\text{AP}}}^{(t)}$ for each time step $t$. By aggregating the information from all APs and through the full time steps $T$, the training dataset $\mathcal{D}^{\text{train}}$ can be formed as: 
	\begin{equation}
		\small
		\begin{split}
			\mathcal{D}^{\text{train}} = \Biggl\{ & \left[ \left[\mathbf{Z}_1^{(0)}, \dots, \mathbf{Z}_{N_{\text{AP}}}^{(0)} \right], \mathbf{L}^{(0)} \right], \left[ \left[\mathbf{Z}_1^{(S)}, \dots, \mathbf{Z}_{N_{\text{AP}}}^{(S)} \right], \mathbf{L}^{(S)} \right], \\ 
			&  \dots, \left[ \left[\mathbf{Z}_1^{(NS)}, \dots, \mathbf{Z}_{N_{\text{AP}}}^{(N  S)} \right], \mathbf{L}^{(N S)} \right] \Biggr\},
		\end{split}
	\end{equation}
	where $S$ is stride, and $N$ can be calculated as follows:
	\begin{equation}
		N = \left\lfloor\frac{T-W}{S}\right\rfloor.
	\end{equation}
	In this case, we introduce a mapping function $f:\mathbb{R}^{N_{\text{\gls{AP}}} \times W \times (2\nu+2)} \rightarrow{\mathbb{R}^{W \times 2}}$, with trainable parameters $\boldsymbol{\Theta}$. Using the function $f$, we map the input observation matrices from all \glspl{AP} into the output predicted locations as:
	\begin{equation}
		\label{eq:map_f}
		\mathbf{L} = f\left(\left[\mathbf{Z}_1, \mathbf{Z}_2, \cdots, \mathbf{Z}_{N_{\text{AP}}}\right]\vert \boldsymbol{\Theta}\right). 
	\end{equation}
	Consequently, the optimal trainable parameters of $f$, denoted by $\boldsymbol{\Theta}^{*}$, can be obtained by minimizing the loss function $\mathcal{L}$ of the predicted and ground-truth locations over $\mathcal{D}^{\text{train}}$ as:
	\begin{equation}
		\label{eq:loss}
		\small
		\boldsymbol{\Theta}^{*} = \argmin_{\boldsymbol{\Theta}} \!\! \sum\limits_{n=0}^N\!\!\mathcal{L}\left(f\left(\left[\mathbf{Z}_1^{(n S)}, \mathbf{Z}_2^{(nS)}, \cdots, \mathbf{Z}_{N_{\text{AP}}}^{(nS)}\right]\vert \boldsymbol{\Theta}\right),\mathbf{L}^{(nS)}\right).
	\end{equation}
	The loss function $\mathcal{L}$ will be discussed in details in next sections. Consequently, during the inference, the optimal parameters $\boldsymbol{\Theta}^*$ is used to predict the output locations as:
	\begin{equation}
		\mathbf{L^{\text{test}}} = f\left(\left[\mathbf{Z}_1^{\text{test}}, \mathbf{Z}_2^{\text{test}}, \cdots, \mathbf{Z}_{N_{\text{AP}}}^{\text{test}}\right] \vert \boldsymbol{\Theta}^*\right).
	\end{equation}
	In Section \ref{sec:methodology}, we discuss our proposed LocAttMamba based solution, representing the mapping function $f$.
	\section{PROPOSED SCHEME}
	\label{sec:methodology}
	This section presents the proposed LocAttMamba framework. First, we review the preliminaries of \glspl{SSM} as the core building block of the Mamba architecture. Subsequently, we introduce the overall architecture of the proposed framework, where each \gls{AP} is first processed using bidirectional Mamba blocks, and then the resulting per-\gls{AP} embeddings are fused using a multi-head attention layer to predict the user location and its per-axis uncertainty.
	\subsection{Preliminaries: SSMs \& Mamba}
	\Glspl{S4} are a recently developed class of deep learning sequence models derived from \glspl{SSM} in control theory, which characterize the evolution of a continuous-time system through its internal state. An \gls{SSM} can be computed either as a recurrence, like a \gls{RNN}, or as a convolution, like a \gls{CNN}, combining efficient inference with parallelizable training \cite{NEURIPS2021_S4}. It performs sequence-to-sequence transformations through an implicit latent state modeled by first-order differential equations. \par 
	As an example, let us denote a continuous input and output sequence $\mathbf{x}(t) \in \mathbb{R}^{N_{\text{in}}}$ and $\mathbf{y}(t)  \in \mathbb{R}^{N_{\text{out}}}$, where $N_{\text{in}}$ and $N_{\text{out}}$ are the number of input features and output variables, respectively. Next, we denote $\mathbf{m}\left(t\right) \in \mathbb{R}^{d_{\text{state}}}$ as the implicit state, with $d_{\text{state}}$ representing the dimension. \glspl{SSM} use three matrices $\mathbf{A} \in \mathbb{R}^{d_{\text{state}} \times d_{\text{state}}}$, $\mathbf{B} \in \mathbb{R}^{ d_{\text{state}} \times N_{\text{in}}}$, and $\mathbf{C} \in \mathbb{R}^{N_{\text{out}} \times d_{\text{state}}}$ as model parameters, which determine the function-to-function transformation through a continuous-time linear system  $\text{\gls{SSM}}\left(\mathbf{A}, \mathbf{B}, \mathbf{C}\right)\left(\mathbf{x}(t)\right)$ as:
	\begin{equation}
		\mathbf{m}'(t) = \mathbf{A}\mathbf{m}(t) + \mathbf{B}\mathbf{x}(t), \quad \quad \mathbf{y}(t) = \mathbf{C} \mathbf{m}(t).
	\end{equation}
	\par 
	To perform sequence-to-sequence transformation, we discretize the continuous parameters of the system using a step size $\delta \in \mathbb{R}_{>0}$. By applying \gls{ZOH}, the  discretized \gls{S4} model $\text{\gls{SSM}}\left(\hat{\mathbf{A}}, \hat{\mathbf{B}}, \hat{\mathbf{C}}, \delta\right)\left(\mathbf{X}\right)$ is expressed as:
	\begin{equation}
		\label{eq: S4_disc}
		\mathbf{m}_k = \hat{\mathbf{A}}\mathbf{m}_{k-1} + \hat{\mathbf{B}}\mathbf{x}_{k}, \quad \mathbf{y}_k = \hat{\mathbf{C}}\mathbf{m}_k.
	\end{equation}
	Here, $\mathbf{m}_k \in \mathbb{R}^{d_{\text{state}}}$, $\mathbf{X} = \left[\mathbf{x}_0, \cdots, \mathbf{x}_k, \cdots, \mathbf{x}_{W-1}\right]^\top \in \mathbb{R}^{W \times N_{\text{in}}}$, $\mathbf{Y} = \left[\mathbf{y}_0, \cdots, \mathbf{y}_k, \cdots, \mathbf{y}_{W-1}\right]^\top \in \mathbb{R}^{W \times N_{\text{out}}}$ are the discretized  state, input, and output, respectively, for a sequence length of $W$, with $\mathbf{x}_k \in \mathbb{R}^{N_{\text{in}}}$ and $\mathbf{y}_k \in \mathbb{R}^{N_{\text{out}}}$ being the $k$-th input and output element. Furthermore, the discretized matrices $\left(\hat{\mathbf{A}}, \hat{\mathbf{B}}, \hat{\mathbf{C}}\right)$ are written based on simplified \gls{ZOH} as:
	
	\begin{align}
		\label{eq:S4_A}
		&\hat{\mathbf{A}} = \exp{\left(\delta\mathbf{A}\right)} \in \mathbb{R}^{d_{\text{state}}\times d_{\text{state}}}, \\
		\label{eq:S4_B}
		&\hat{\mathbf{B}} = \left(\delta\mathbf{A}\right)^{-1}\left(\exp{\left(\delta\mathbf{A}\right)}-\mathbf{I}_{d_{\text{state}}}\right)\left(\delta\mathbf{B}\right) \in \mathbb{R}^{d_{\text{state}} \times N_{\text{in}}}, \\
		\label{eq:S4_C}
		&\hat{\mathbf{C}} = \mathbf{C} \in \mathbb{R}^{N_{\text{out}}\times d_{\text{state}}}. 
	\end{align}
	\par 
	In the \gls{S4} model shown in (\ref{eq: S4_disc})-(\ref{eq:S4_C}), matrices $\left(\hat{\mathbf{A}}, \hat{\mathbf{B}}, \hat{\mathbf{C}}, \delta\right)$ are assumed to be constant, enabling a \gls{LTI} system. In particular, the \gls{LTI} feature enables the computation of these models through a global convolution \cite{NEURIPS2021_S4}, significantly reducing the inference. However, the input independent structure of this \gls{SSM} limits its ability in more complex problems. \par 
	\begin{figure}[t!]
		\centering
		\includegraphics[trim = 20 73 20 38, clip, width=0.7\linewidth]{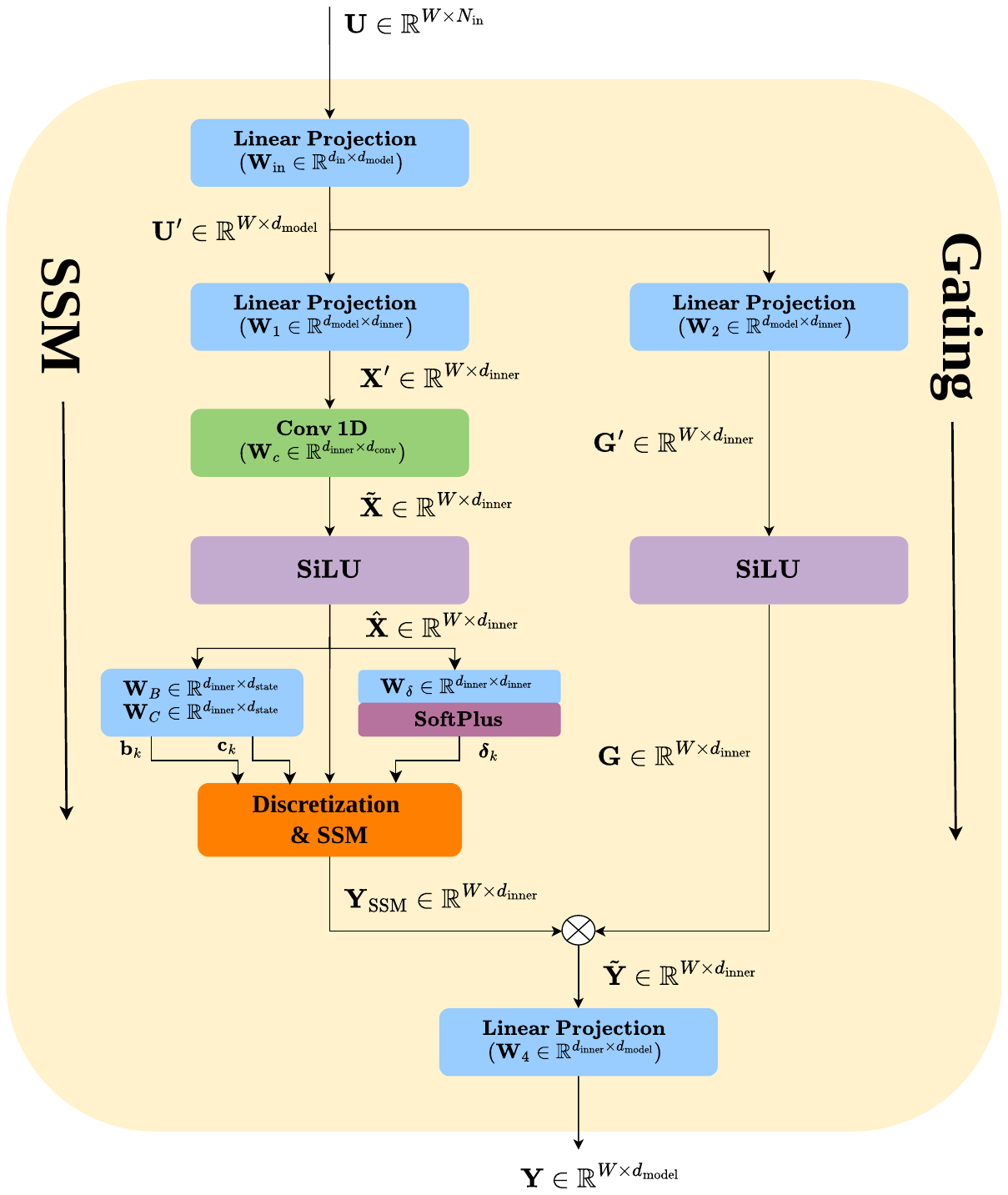}
		\caption{Block diagram of the Mamba block.}
		\label{fig:Mamba_Block}
	\end{figure}
	\subsection{Mamba Block}
	While the S4 model in Section III-A is efficient, its input-independent parameters limit its expressiveness. Mamba overcomes this by introducing a selection mechanism, in which the matrices $\mathbf{B}, \mathbf{C}$, and $\delta$ become functions of the input \cite{gu2023mamba}. A scalar \gls{SSM} is applied independently to each input channel, so that the discretized parameters, and hence the recurrence, become time-varying. We first describe the block that produces these input-dependent parameters, and then the resulting selective recurrence. \par 
	Fig. \ref{fig:Mamba_Block} shows the block diagram of the Mamba block \cite{gu2023mamba}. Let $\mathbf{U} \in \mathbb{R}^{W \times N_{\text{in}}}$ be the block input. The input is projected into dimension $d_{\text{model}}$ using weight $\mathbf{W}_{\text{in}} \in \mathbb{R}^{N_{\text{in}} \times d_{\text{model}}}$ as:
	\begin{equation}
		\mathbf{U}' = \mathbf{U} \mathbf{W}_{\text{in}} \in \mathbb{R}^{W \times d_{\text{model}}}.
	\end{equation}
	
    Using $\mathbf{U}'$, two branches are created. In the \gls{SSM} branch, a linear projection is applied to $\mathbf{U}'$ as:
	\begin{equation}
		\mathbf{X}' = \mathbf{U}'\mathbf{W}_1 \in \mathbb{R}^{W \times d_{\text{inner}}},
	\end{equation} 
	where $\mathbf{W}_1 \in \mathbb{R}^{d_{\text{model}} \times d_{\text{inner}}}$ is the projection weight, and $d_{\text{inner}}$ is the dimension to which the input is projected. \\
	
    The output $\mathbf{X}'$ is passed through depthwise \gls{1-D} Convolution with weight $\mathbf{W}_c \in \mathbb{R}^{d_{\text{inner}} \times d_{\text{conv}}}$ as: 
	\begin{equation}
		\hat{\mathbf{X}} = \mathrm{Conv1D}(\mathbf{X}', \mathbf{W}_c)\in \mathbb{R}^{W \times d_{\text{inner}}},
	\end{equation}
	which is then passed through a SiLU function: 
	\begin{equation}
		\tilde{\mathbf{X}} = \mathrm{SiLU}\left(\hat{\mathbf{X}}\right) \in \mathbb{R}^{W \times d_{\text{inner}}},
	\end{equation}
	where $\mathrm{SiLU}(z) = z \cdot \frac{1}{1 + e^{-z}}$. \\ 
	Subsequently, rows of the obtained
	\begin{equation}
		\tilde{\mathbf{X}} = \left[\tilde{\mathbf{x}}_0, \cdots, \tilde{\mathbf{x}}_k, \cdots, \tilde{\mathbf{x}}_{W-1}\right]^\top \in \mathbb{R}^{W \times d_{\text{inner}}},   
	\end{equation}
	with $\tilde{\mathbf{x}}_k \in \mathbb{R}^{d_{\text{inner}}}$ being the $k$-th element, are used to form input-dependent vectors $\mathbf{b}_k \in \mathbb{R}^{ d_{\text{state}}}$, $\mathbf{c}_k \in \mathbb{R}^{ d_{\text{state}}}$, and $\boldsymbol{\delta}_k \in \mathbb{R}^{d_{\text{inner}}}$ as:
	\begin{equation}
		\begin{split}
			\label{eq:mamba_main}
			\mathcal{\mathbf{b}}_k = & \tilde{\mathbf{x}}_k\mathbf{W}_B \in \mathbb{R}^{ d_{\text{state}}}, \quad \quad
			\mathbf{c}_k = \tilde{\mathbf{x}}_k\mathbf{W}_C\in \mathbb{R}^{ d_{\text{state}}}, \\ \quad
			& \quad  \ \boldsymbol{\delta}_k = \mathrm{softplus}\!\left(\tilde{\mathbf{x}}_k\mathbf{W}_{\delta}\right) \in \mathbb{R}^{d_{\text{inner}}},
		\end{split}
	\end{equation}
	where $\mathbf{W}_B, \mathbf{W}_C \in \mathbb{R}^{d_{\text{inner}} \times d_{\text{state}}}$
	are the projection weights of the state matrices, and $\mathbf{W}_{\delta} \in \mathbb{R}^{d_{\text{inner}} \times d_{\text{inner}}}$ is the projection weight for the time step. This yields $\mathbf{b}_k, \mathbf{c}_k \in \mathbb{R}^{d_{\text{state}}}$
	and $\boldsymbol{\delta}_k \in \mathbb{R}^{d_{\text{inner}}}$, and the
	$\mathrm{softplus}$ function is defined as:
	\begin{equation}
		\mathrm{softplus}(z) = \log\left(1 + e^{z}\right),
	\end{equation}
	which guarantees $\boldsymbol{\delta}_k > 0$. \par
	Since a scalar \gls{SSM} is applied per channel, the state matrix $\mathbf{A} \in \mathbb{R}^{d_{\text{inner}} \times d_{\text{state}}}$ is
	diagonal and stored per channel,  where each row contains the diagonal entries of one channel. The \gls{ZOH}
	in (\ref{eq:S4_A})--(\ref{eq:S4_C}) is then applied per time step,
	so that the discretized matrices become input-dependent:
	\begin{align}
		\label{eq:mamba_disc}
		& \hat{\mathbf{A}}_k = \exp\!\left(\text{diag}\left(\boldsymbol{\delta}_k \right)\mathbf{A}\right) \in \mathbb{R}^{d_{\text{inner}}\times d_{\text{state}}}, \\
		& \hat{\mathbf{B}}_k = \left(\boldsymbol{\delta}_k \odot \tilde{\mathbf{x}}_k\right)\mathbf{b}_k^\top \in \mathbb{R}^{d_{\text{inner}}\times d_{\text{state}}} , \\
		& \hat{\mathbf{c}}_k = \mathbf{c}_k \in \mathbb{R}^{d_{\text{state}}},
	\end{align}
	where $\hat{\mathbf{A}}_k, \hat{\mathbf{B}}_k \in \mathbb{R}^{d_{\text{inner}} \times d_{\text{state}}}$
	and $\odot$ denotes element-wise multiplication with broadcasting over the
	state dimension. Consequently, the recurrence in (\ref{eq: S4_disc})
	becomes time-varying for Mamba as:
	\begin{equation}
		\label{eq:mamba_rec}
		\mathbf{M}_k = \hat{\mathbf{A}}_k \odot \mathbf{M}_{k-1} + \hat{\mathbf{B}}_k, \qquad
		\mathbf{y}_k = \mathbf{M}_k\hat{\mathbf{c}}_k ,
	\end{equation}
	where $\mathbf{M}_k \in \mathbb{R}^{d_{\text{inner}} \times d_{\text{state}}}$
	is the state matrix, $\mathbf{y}_k \in \mathbb{R}^{d_{\text{inner}}}$ is
	the output. Here, $\mathbf{A}$ remains a static learned parameter and
	the time-variation is introduced solely through $\boldsymbol{\delta}_k$,
	$\mathbf{B}_k$, and $\mathbf{c}_k$. Notably, the state converts to a matrix from a vector, due to the fact that Mamba considers independent \glspl{SSM} for each of the input features, resulting in having $d_{\text{inner}}$ independent state vectors, and thus, having a $d_{\text{inner}} \times d_{\text{state}}$ state matrix at each time step. The full \gls{SSM} output is then written as:
	\begin{equation}
		\mathbf{Y}_{\text{\gls{SSM}}} = \left[\mathbf{y}_0, \cdots, \mathbf{y}_k, \cdots, \mathbf{y}_{W-1}\right]^\top \in \mathbb{R}^{W \times d_{\text{inner}}}.
	\end{equation}
	\par 
	In the gating branch, the $\mathbf{U}'$ matrix is first projected using the weight $\mathbf{W}_2 \in \mathbb{R}^{d_{\text{model}} \times d_{\text{inner}}}$ as:
	\begin{equation}
		\mathbf{G}' = \mathbf{U}'\mathbf{W}_2 \in \mathbb{R}^{W \times d_{\text{inner}}},
	\end{equation}
	which is then passed through a $\mathrm{SiLU}$ function:
	\begin{equation}
		\mathbf{G} = \mathrm{SiLU}(\mathbf{G}') \in \mathbb{R}^{W \times d_{\text{inner}}}.
	\end{equation}
	Finally, the outputs from the SSM and gating branch are combined using element-wise multiplication as:
	\begin{equation}
		\tilde{\mathbf{Y}} = \mathbf{Y}_{\text{SSM}} \odot \mathbf{G} \in \mathbb{R}^{W \times d_{\text{inner}}},
	\end{equation} 
	which is passed through the linear weight $\mathbf{W}_4 \in \mathbb{R}^{d_{\text{inner}} \times d_{\text{model}}}$ to obtain the Mamba model output:
	\begin{equation}
		\mathbf{Y} = \tilde{\mathbf{Y}} \mathbf{W}_4 \in \mathbb{R}^{W \times d_{\text{model}}}.
	\end{equation}
	\par It is important to note that since Mamba uses input-dependent model parameters, the resulting system is no longer \gls{LTI}, and thus, it cannot be evaluated as a global convolution; instead, Mamba employs a hardware-aware selective scan to compute (\ref{eq:mamba_rec}) efficiently instead of the normal sequential one  \cite{gu2023mamba}. We will demonstrate the impact of using parallel scan in the next section. 
    
	\subsection{Proposed LocAttMamba Framework}
	In this section, we will explain the proposed architecture of the mapping function $f\left(.\right)$ for the user localization problem. \par
\begin{figure*}[!t]
		\centering
		\includegraphics[trim= 20 495 25 22, clip, width=0.8\textwidth]{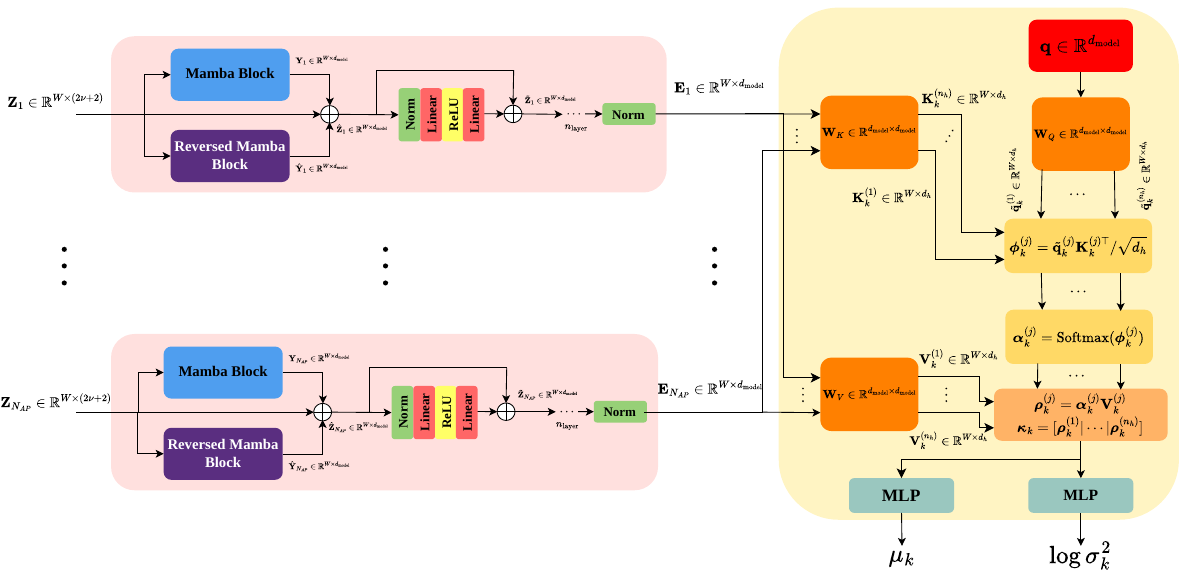}
		\caption{Block diagram of the proposed LocAttMamba approach.}
		\label{fig:Mamba_Attention}
	\end{figure*}

	Fig. \ref{fig:Mamba_Attention} illustrates the proposed architecture. Specifically, we consider $N_{\text{\gls{AP}}}$ encoders, each receiving an input sequence $\mathbf{Z}_i^{(nS)} \in \mathbb{R}^{W \times (2\nu+2)}$ from the training dataset $\mathcal{D}^{\mathrm{train}}$, where $n \in \left\{0, \ldots, N\right\}$. For notational simplicity, we omit the superscript $(nS)$ in the following discussion. Each input sequence is then processed by its corresponding encoder through a stack of Mamba blocks, shown in Fig. \ref{fig:Mamba_Block}, as follows:
	\begin{equation}
		\begin{split}
			& \mathbf{Y}_i = \mathrm{Mamba}\left(\mathbf{Z}_i\right) \in \mathbb{R}^{W \times d_{\text{model}}}, \\
			& \hat{\mathbf{Y}}_i = \mathbf{J} \ \mathrm{Mamba}\left(\mathbf{J}\mathbf{Z}_i\right) \in \mathbb{R}^{W \times d_{\text{model}}},
		\end{split}
	\end{equation}
	where $\mathbf{J} \in \mathbb{R}^{W \times W}$ is an anti-diagonal matrix with the elements along the anti-diagonal being 1, and all other elements being 0, forming the reverse Mamba. 
	Projecting the input into $d_{\text{model}}$ and adding that as a residual branch, the outputs are added up to obtain $\hat{\mathbf{Z}}_i \in \mathbb{R}^{W \times d_{\text{model}}}$ as:
	\begin{equation}
		\hat{\mathbf{Z}}_i = \mathbf{Y}_i + \hat{\mathbf{Y}}_i + \mathbf{Z}_i\mathbf{W}_{\text{res}} \in \mathbb{R}^{W \times d_{\text{model}}}, 
	\end{equation}
    \par
	 where $\mathbf{W}_{\text{res}} \in \mathbb{R}^{(2\nu+2)\times d_{\text{model}}}$ is the input projection weight. Subsequently, $\hat{\mathbf{Z}}_i$ is passed through several linear projection layers as:
	\begin{equation}
		\tilde{\mathbf{Z}}_i = \hat{\mathbf{Z}}_i +  \mathrm{ReLU}\left(\mathrm{Norm}\left(\hat{\mathbf{Z}}_i\right)\mathbf{W}_{\psi_1}\right)\mathbf{W}_{\psi_2} \in \mathbb{R}^{W \times d_{\text{model}}},
	\end{equation}
	where $\tilde{\mathbf{Z}}_i \in \mathbb{R}^{W \times d_{\text{model}}}$ is the output, $\mathbf{W}_{\psi_1} \in \mathbb{R}^{d_{\text{model}} \times d_{\text{model}}}$ and $\mathbf{W}_{\psi_2} \in \mathbb{R}^{d_{\text{model}} \times d_{\text{model}}}$ are the projection weights, the $\mathrm{ReLU}$ function is defined as:
	\begin{equation}
		\begin{split}
			& \mathrm{ReLU}(z) = \max\left(0, z\right),
		\end{split}
	\end{equation}
	and $\mathrm{Norm}$ represents the root mean square normalization which can be written for a vector $\mathbf{u} \in \mathbb{R}^{d}$ as:
	\begin{equation}
		\mathrm{Norm}\left(\mathbf{u}\right) = \boldsymbol{\gamma} \odot \frac{\mathbf{u}}{\sqrt{\frac{1}{d}\sum\limits_i u_i^2}},
	\end{equation}
	where $\boldsymbol{\gamma} \in \mathbb{R}^{d_{\text{model}}}$ is a vector with $d_{\text{model}}$ trainable elements, for each $\mathrm{Norm}$ layer. \\
	This sequence is repeated for $n_{\text{layer}}$, and the output is passed through a $\mathrm{Norm}$ layer to obtain the embedding $\mathbf{E}_i \in \mathbb{R}^{W \times d_{\text{model}}}$.
    \par 
	Subsequently, first, the embeddings $\mathbf{E}_i$ are stacked into a single tensor $\boldsymbol{\mathcal{E}} \in \mathbb{R}^{N_{\text{\gls{AP}}} \times W \times d_{\text{model}}}$ as:
	\begin{equation}
		\begin{split}
			& \boldsymbol{\mathcal{E}} = \mathrm{stack}(\mathbf{E}_1,\dots,\mathbf{E}_{N_{\text{\gls{AP}}}}) \in \mathbb{R}^{ N_{\text{\gls{AP}}}\times W \times d_{\text{model}}}, \\ & \boldsymbol{\mathcal{E}}_k \in \mathbb{R}^{N_{\text{\gls{AP}}}\times d_{\text{model}}} \ \ \ \forall k \in \left\{0, \cdots, W-1\right\},
		\end{split}
	\end{equation}	
	where $\boldsymbol{\mathcal{E}}_k \in \mathbb{R}^{N_{\text{\gls{AP}}}\times d_{\text{model}}}$ denotes the stacked embedding for the $k$-th time step within the window $W$. \par 
	To this point, we have obtained the output embeddings of the Mamba blocks and stacked them
	into $\boldsymbol{\mathcal{E}}_k$. To combine these embeddings across the
	$N_{\text{\gls{AP}}}$ \glspl{AP} and obtain an accurate track estimate, we use an attention
	mechanism \cite{Vaswani2017Attention, pmlr-v80-ilse18a} that weights each \gls{AP} according to its importance at every time step. To do
	so, the learnable projections
	$\mathbf{W}_Q,\mathbf{W}_K,\mathbf{W}_V\in\mathbb{R}^{d_{\text{model}}\times d_{\text{model}}}$
	map a shared query $\mathbf{q} \in \mathbb{R}^{d_{\text{model}}}$ and the stacked embedding $\boldsymbol{\mathcal{E}}_k$ into
	the query $\hat{\mathbf{q}} \in \mathbb{R}^{d_{\text{model}}}$, key $\mathbf{K}_k \in \mathbb{R}^{N_{\text{\gls{AP}}}\times d_{\text{model}}}$, and value $\mathbf{V}_k \in \mathbb{R}^{N_{\text{\gls{AP}}}\times d_{\text{model}}}$ at each time step $k$:
	\begin{align}
		& \hat{\mathbf{q}} = \mathbf{q}\mathbf{W}_Q^\top \in \mathbb{R}^{d_{\text{model}}}, \\
		& \mathbf{K}_k = \boldsymbol{\mathcal{E}}_k \mathbf{W}_K^\top \in \mathbb{R}^{N_{\text{\gls{AP}}} \times d_{\text{model}}},\\
		& \mathbf{V}_k = \boldsymbol{\mathcal{E}}_k \mathbf{W}_V^\top \in \mathbb{R}^{N_{\text{\gls{AP}}} \times d_{\text{model}}}.
	\end{align}
    \par
	For multi-head attention, these matrices are split into $n_H$ heads, where each head
	$j\in\{1,\dots,n_H\}$ works on a subspace of dimension
	$d_h = d_{\text{model}}/n_H$ (assuming $d_{\text{model}}$ is divisible by $n_H$).
	Splitting $\hat{\mathbf{q}}$, $\mathbf{K}_k$, and $\mathbf{V}_k$ along their feature dimension gives the per-head slices as:
	\begin{align}
		\hat{\mathbf{q}}^{(j)} \in \mathbb{R}^{d_h}, \quad
		\mathbf{K}_k^{(j)} \in \mathbb{R}^{N_{\text{\gls{AP}}} \times d_h}, \quad
		\mathbf{V}_k^{(j)} \in \mathbb{R}^{N_{\text{\gls{AP}}} \times d_h}.
	\end{align}
	Subsequently, we compute the attention score $\boldsymbol{\phi}_k^{(j)} \in \mathbb{R}^{N_{\text{\gls{AP}}}}$ for each head as follows:
	\begin{equation}
		\boldsymbol{\phi}_k^{(j)} = \frac{\hat{\mathbf{q}}^{(j)}\big(\mathbf{K}_k^{(j)}\big)^\top}{\sqrt{d_h}}
		\in \mathbb{R}^{N_{\text{\gls{AP}}}},
	\end{equation}
	which is normalized using $\mathrm{Softmax}$ function to obtain normalized attention weights $\boldsymbol{\alpha}_k^{(j)} \in \mathbb{R}^{N_{\text{\gls{AP}}}}$ as:
	\begin{equation}
		\boldsymbol{\alpha}_k^{(j)} = \mathrm{Softmax}\big(\boldsymbol{\phi}_k^{(j)}\big) \in \mathbb{R}^{N_{\text{\gls{AP}}}}, \ \ \ \mathrm{Softmax}(\mathbf{z})_p = \frac{e^{z_p}}{\sum\limits_pe^{z_p}},
	\end{equation}
	which is used to compute the context vector $\boldsymbol{\rho}_k^{(j)} \in \mathbb{R}^{d_h}$ as:
	\begin{equation}
		\boldsymbol{\rho}_k^{(j)}= \boldsymbol{\alpha}_k^{(j)}\mathbf{V}_k^{(j)} \in \mathbb{R}^{d_h}. 
	\end{equation}
	Next, the obtained context vectors are concatenated and projected using projection weight $\mathbf{W}_o \in \mathbb{R}^{d_{\text{model}} \times d_{\text{model}}}$ as:
	\begin{equation}
		\boldsymbol{\kappa}_k = \big[\boldsymbol{\rho}_k^{(1)} \,\Vert\, \cdots \,\Vert\, \boldsymbol{\rho}_k^{(n_H)}\big]\mathbf{W}_o^\top
		\in \mathbb{R}^{d_{\text{model}}}, \ \mathbf{W}_o\in\mathbb{R}^{d_{\text{model}}\times d_{\text{model}}}.
	\end{equation}
	Notably, the contribution of each \gls{AP}'s embeddings in the fusion is determined by the average normalized attention weight $\bar{\boldsymbol{\alpha}}_k \in \mathbb{R}^{N_{\text{\gls{AP}}}}$ across attention heads as:
	\begin{equation}
		\label{eq:attention_weight}
		\bar{\boldsymbol{\alpha}}_k = \frac{1}{n_H}\sum_{j=1}^{n_H}\boldsymbol{\alpha}_k^{(j)} \in \mathbb{R}^{N_{\text{\gls{AP}}}}.
	\end{equation}
	Consequently, the predicted \gls{2-D} location $\boldsymbol{\mu}_k = \left[\mu_{k, x}, \mu_{k,y}\right] \in \mathbb{R}^2$ and uncertainty $\boldsymbol{\sigma}_k  \in \mathbb{R}^2$ are obtained as:
	\begin{align}
		\boldsymbol{\mu}_k &=  \left[\mu_{k, x}, \mu_{k,y}\right]^\top = \mathbf{W}_2^{\mu} \, \mathrm{SiLU}\left(\mathbf{W}_1^{\mu}\boldsymbol{\kappa}_k\right) \in \mathbb{R}^{2}, \\
		\log\boldsymbol{\sigma}_k^2 &= \left[\log\sigma_{k, x}^2, \log\sigma_{k,y}^2\right] ^\top=  \mathbf{W}_2^{\sigma}\,\mathrm{SiLU}\left(\mathbf{W}_1^{\sigma}\boldsymbol{\kappa}_k\right)\in \mathbb{R}^{2}, 
	\end{align}
	where $\mu_{k,x} \in \mathbb{R}$ and $\mu_{k,y} \in \mathbb{R}$ denote the predicted location along the $x$- and $y$-axes, and $\mathbf{W}_1^{\mu} \in \mathbb{R}^{d_{\text{model}} \times d_{\text{model}}}$ and $\mathbf{W}_2^{\mu} \in \mathbb{R}^{2 \times d_{\text{model}}}$ are the first and second projection weights of the location head. Similarly, $\sigma_{k,x} \in \mathbb{R}$ and $\sigma_{k,y} \in \mathbb{R}$ denote the predicted uncertainty along the two axes, with $\mathbf{W}_1^{\sigma} \in \mathbb{R}^{d_{\text{model}} \times d_{\text{model}}}$ and $\mathbf{W}_2^{\sigma} \in \mathbb{R}^{2 \times d_{\text{model}}}$ the corresponding projection weights. Eventually, by stacking the predicted location and uncertainty over all time steps, we obtain the following:
	\begin{equation}
		\boldsymbol{\mu}\in\mathbb{R}^{W\times 2}, \qquad
		\log\boldsymbol{\sigma}^2\in\mathbb{R}^{W\times 2}. 
	\end{equation}
	The discussed architecture works as the mapping function $f\left(.\right)$ written in (\ref{eq:map_f}), with $\boldsymbol{\Theta}$ representing the parameters of the proposed LocAttMamba. The summary of this approach is shown in Algorithm \ref{alg:mamba_attention}. 
	\subsection{Loss Functions}
	To optimize the trainable parameters $\boldsymbol{\Theta}$ by minimizing the loss in (\ref{eq:loss}), we first use \gls{MSE} loss as: 
	\begin{equation}
		\label{eq:mse}
		\mathcal{L}_{\text{MSE}} = \frac{1}{2W}\sum_{k=0}^{W-1} \left\| \boldsymbol{\mu}_k - \mathbf{l}_k \right\|_2^2,
	\end{equation}
	which is used during the warm-up training. After warm-up, the model parameters are optimized using $\beta$-\gls{NLL} loss as \cite{seitzer2022on}:
	\begin{equation}
		\label{eq:beta_nll}
		\mathcal{L}_{\beta\text{-NLL}} = \frac{1}{2W}\sum_{k=0}^{W-1} \left\lfloor \sigma_k^2 \right\rfloor^{\beta} \left( \log \sigma_k^2 + \frac{\left(\mathbf{l}_k - \boldsymbol{\mu}_k\right)^2}{\sigma_k^2} \right),
	\end{equation}
	which is used to simultaneously optimize $\boldsymbol{\mu}$ and uncertainty head $\log\boldsymbol{\sigma}$, where $\left\lfloor.\right\rfloor$ denotes the stop-gradient operator \cite{seitzer2022on}.

    \begin{algorithm}[!t]
	\caption{Proposed LocAttMamba algorithm}
	\label{alg:mamba_attention}
	\SetAlgoLined
	\DontPrintSemicolon
	\KwIn{Per-\gls{AP} feature sequences $\{\mathbf{Z}_i\}_{i=1}^{N_{\text{\gls{AP}}}}$, $\mathbf{Z}_i \in \mathbb{R}^{W \times (2\nu+2)}$.}
	\KwOut{Predicted locations $\boldsymbol{\mu} \in \mathbb{R}^{W \times 2}$ and per-axis log-variances $\log\boldsymbol{\sigma}^2 \in \mathbb{R}^{W \times 2}$.}
	\BlankLine
	\For{each \gls{AP} $i = 1, \dots, N_{\text{\gls{AP}}}$}{
		\For{each encoder layer $\ell = 1, \dots, n_{\text{layer}}$}{
			Process $\mathbf{Z}_i$ in forward temporal direction $\mathbf{Y}_i \leftarrow \mathrm{Mamba}(\mathbf{Z}_i)$\;
			Process $\mathbf{Z}_i$ in reverse temporal direction $\hat{\mathbf{Y}}_i \leftarrow \mathbf{J}\,\mathrm{Mamba}(\mathbf{J}\mathbf{Z}_i)$\;
			Merge both directions through a residual connection:
			$\hat{\mathbf{Z}}_i \leftarrow \mathbf{Y}_i + \hat{\mathbf{Y}}_i + \mathbf{Z}_i\mathbf{W}_{\text{res}}$\;
			Update $\mathbf{Z}_i$ via a normalized feed-forward block with residual connection\;
		}
    \gls{AP} embedding: $\mathbf{E}_i \leftarrow \mathrm{Norm}(\mathbf{Z}_i)$\;
	}
	Stack the per-\gls{AP} embeddings into $\boldsymbol{\mathcal{E}} \leftarrow \mathrm{stack}(\mathbf{E}_1, \dots, \mathbf{E}_{N_{\text{\gls{AP}}}})$\;
	Project the shared learnable query: $\hat{\mathbf{q}} \leftarrow \mathbf{q}\mathbf{W}_Q^\top$\;
	\BlankLine
	\For{each time step $k = 0, \dots, W-1$}{
		Project the \gls{AP} embeddings at time step $k$ into keys and values:
		$\mathbf{K}_k \leftarrow \boldsymbol{\mathcal{E}}_k\mathbf{W}_K^\top$,\;
		$\mathbf{V}_k \leftarrow \boldsymbol{\mathcal{E}}_k\mathbf{W}_V^\top$\;
		\For{each attention head $j = 1, \dots, n_H$}{
			Compute the importance weight of each \gls{AP}:
			$\boldsymbol{\alpha}_k^{(j)} \leftarrow \mathrm{Softmax}\big(\hat{\mathbf{q}}^{(j)}(\mathbf{K}_k^{(j)})^\top / \sqrt{d_h}\big)$\;
			Aggregate the \gls{AP} values: $\boldsymbol{\rho}_k^{(j)} \leftarrow \boldsymbol{\alpha}_k^{(j)}\mathbf{V}_k^{(j)}$\;
		}
		Concatenate and project heads to obtain $\boldsymbol{\kappa}_k$\;
		Predict $\boldsymbol{\mu}_k$ and $\log\boldsymbol{\sigma}_k^2$ from $\boldsymbol{\kappa}_k$ via separate heads\;
	}
	Collect the predictions over all time steps into $\boldsymbol{\mu}$ and $\log\boldsymbol{\sigma}^2$\;
	\Return{$\boldsymbol{\mu}$,\ $\log\boldsymbol{\sigma}^2$}
\end{algorithm}
	\subsection{Post-hoc Uncertainty Calibration}
	Since models trained with the $\beta$-NLL objective tend to be
	overconfident, we apply a post-hoc calibration step to the predicted
	per-axis variance. Let $\mathbf{e}_k = \boldsymbol{\mu}_k - \mathbf{l}_k$
	denote the prediction error at time step $k$ and
	$(\sigma_{k,x}, \sigma_{k,y})$ the predicted standard deviations. The
	calibration is measured by the mean normalized squared error:
	\begin{equation}
		\label{eq:chi2}
		\chi^2 = \frac{1}{N}\sum_{k=1}^{N}\left(\frac{e_{k,x}^2}{\sigma_{k,x}^2} + \frac{e_{k,y}^2}{\sigma_{k,y}^2}\right),
	\end{equation}
	where $\chi^2=2$ indicates a well-calibrated two-dimensional Gaussian, with
	larger values indicating overconfidence. To correct any miscalibration,
	we apply per-axis temperature scaling \cite{guo2017calibration}, with
	scales fit on the validation set as:
	\begin{equation}
		\label{eq:temp_scale}
		s_x = \sqrt{\frac{1}{N_{\text{val}}}\sum_{k}\frac{e_{k,x}^2}{\sigma_{k,x}^2}}, \quad
		s_y = \sqrt{\frac{1}{N_{\text{val}}}\sum_{k}\frac{e_{k,y}^2}{\sigma_{k,y}^2}},
	\end{equation}
	yielding calibrated uncertainties $\tilde{\sigma}_{k,x} = s_x\,\sigma_{k,x}$
	and $\tilde{\sigma}_{k,y} = s_y\,\sigma_{k,y}$, which are applied unchanged
	to the test set. The closed form in (\ref{eq:temp_scale}) is the exact
	minimizer of the Gaussian negative log-likelihood over a per-axis scale.
	\section{NUMERICAL RESULTS}
	\label{sec:NUM}
	\begin{table}[!t]
		\centering
		\caption{Model and Training Hyper-parameters.}
		\label{tab:mamba_hyperparams}
        \resizebox{\linewidth}{!}{%
		\begin{tabular}{lclc}
			\hline
			\textbf{Hyperparameter} & \textbf{Value} & \textbf{Hyperparameter} & \textbf{Value} \\
			\hline
			Number of Layers ($n_{\text{layer}}$) & 2  & Learning Rate ($\eta$) & 0.003 \\
			State Dimension ($d_{\text{state}}$)  & 16 & Batch Size ($\mathcal{B}_S$)             & 16    \\
			Inner Dimension ($d_{\text{inner}}$)  & 32 & Model Dimension ($d_{\text{model}}$)  & 32   \\
			Attention heads ($n_H$)  & 4 & Weight Decay $(\lambda)$           & 0.05  \\
			Window Size $(W)$                       & 25 & Stride $(S)$                & 5   \\
			Warm-up Epochs                       & 6 & Total Epochs                & 15   \\
			$\beta$                       & 0.5 & Optimizer                & AdamW   \\
			\hline
		\end{tabular}}
	\end{table}
	In this section, we present the numerical results obtained from our simulations. We investigate a real-world \gls{5G} indoor localization dataset in \cite{fraunhofer_dataset}, which includes $N_{\text{\gls{AP}, \gls{5G}}} = 6$ \glspl{AP} and $\nu_{\text{\gls{5G}}} = 128$ \gls{CIR} elements in addition to \gls{TDoA} and \gls{ToA}-offset features. The dataset is publicly available and was collected with a bandwidth of $100$ MHz and a carrier frequency of $3.7$ GHz. Notably, the dataset provides separate train and test sets with $T_{\text{train,\gls{5G}}} = 18867$ and $T_{\text{test,\gls{5G}}} = 15722$, where the test set is held out until the final evaluation. From the training set, the last $20\%$ of time steps are used as the validation set, where the split is applied before windowing so that no window spans the train–validation boundary. \par 
	To improve the reliability of the results, all simulations are conducted over 4 independent runs, and the reported performance metrics are averaged across these runs. Furthermore, we consider validation distance as the key evaluation metric defined as:
	\begin{equation}
		\label{eq:val_dist}
		d =  \sqrt{\left(l_x - \mu_x\right)^2 + \left(l_y - \mu_y\right)^2},
	\end{equation}
	where $l_x$ $(l_y)$ and $\mu_x$ $(\mu_y)$ are the ground-truth and predicted locations along the x-axis (y-axis), respectively. \par 
	We conduct experiments using the Kaggle computing platform with an Intel Xeon CPU running at approximately 2.2 GHz and two NVIDIA Tesla T4 GPUs with 16 GB memory each. For training, we use AdamW optimizer \cite{loshchilov2019decoupled} with weight decay parameter $\lambda$ for 15 epochs in total and 6 warm-up epochs. The input features are first normalized using Z-score normalization, where both the validation and test set use the statistics obtained from the training set for later normalization.  Moreover, we use window size $W=25$ and stride $S=5$ for all the results presented in this paper.  
	\begin{figure*}[htbp]
		\centering
		\subfloat[Learning rate ($\eta$)]{%
			\includegraphics[width=0.23\textwidth]{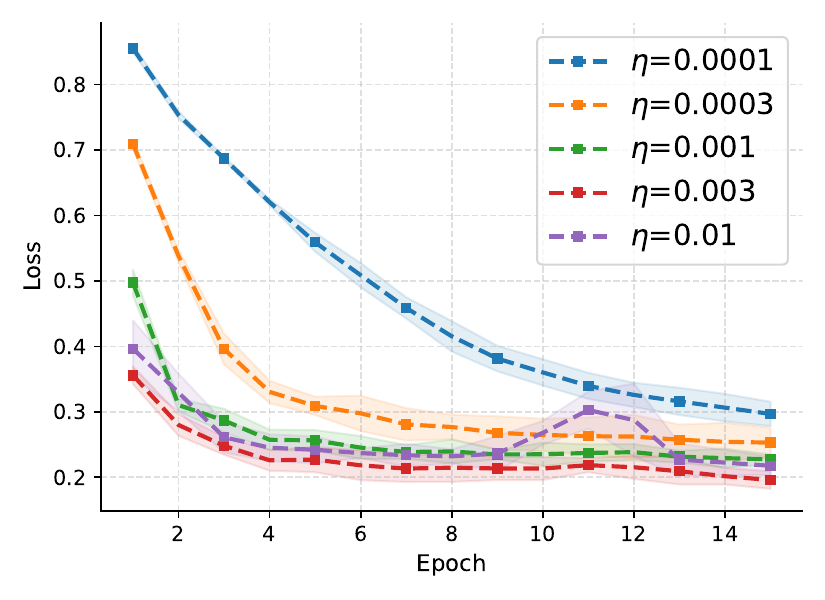}%
			\includegraphics[width=0.23\textwidth]{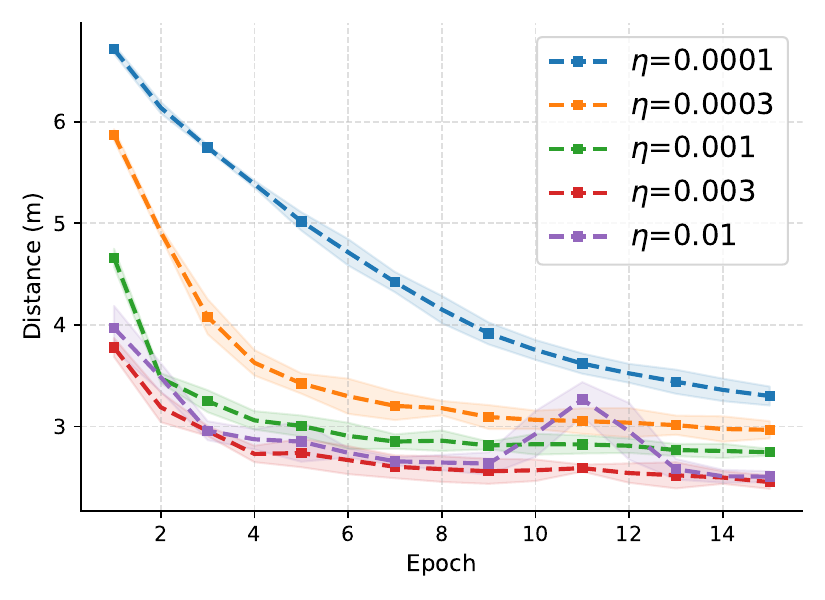}%
			\label{fig:Hyper_Tuneg1}%
		}
		\hfil
		\subfloat[Batch size ($\mathcal{B}_S$)]{%
			\includegraphics[width=0.23\textwidth]{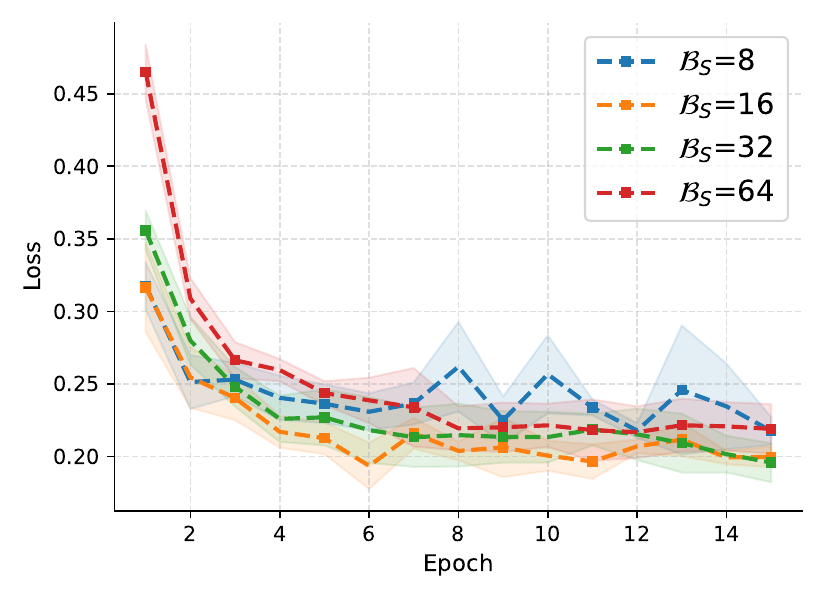}%
			\includegraphics[width=0.23\textwidth]{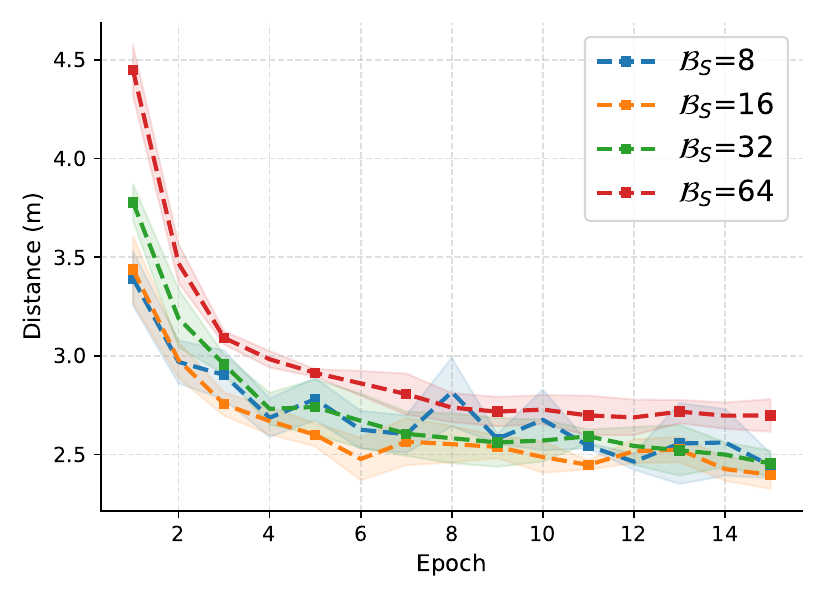}%
			\label{fig:Hyper_Tuneg2}%
		}
		
		
		\subfloat[State dimension ($d_{\text{state}}$)]{%
			\includegraphics[width=0.23\textwidth]{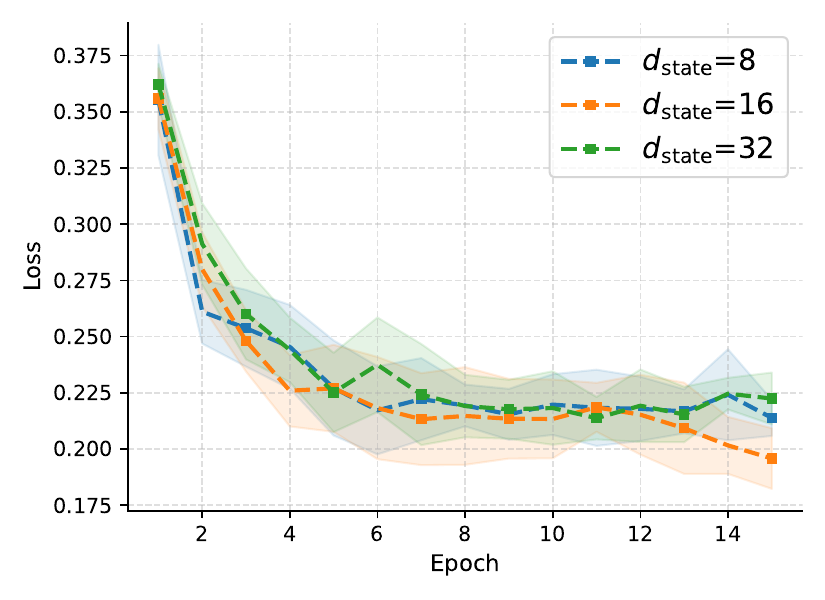}%
			\includegraphics[width=0.23\textwidth]{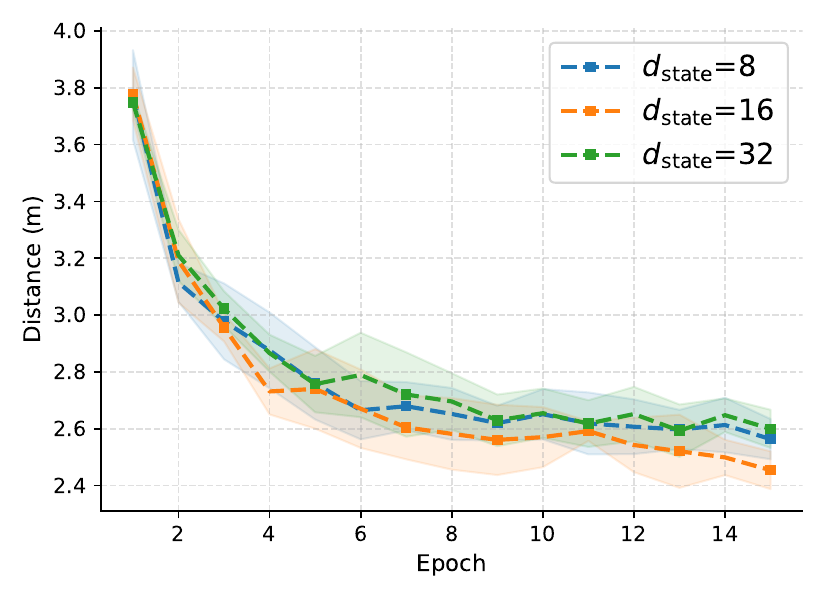}%
			\label{fig:Hyper_Tuneg3}%
		}
		\hfil
		\subfloat[Weight decay ($\lambda$)]{%
			\includegraphics[width=0.23\textwidth]{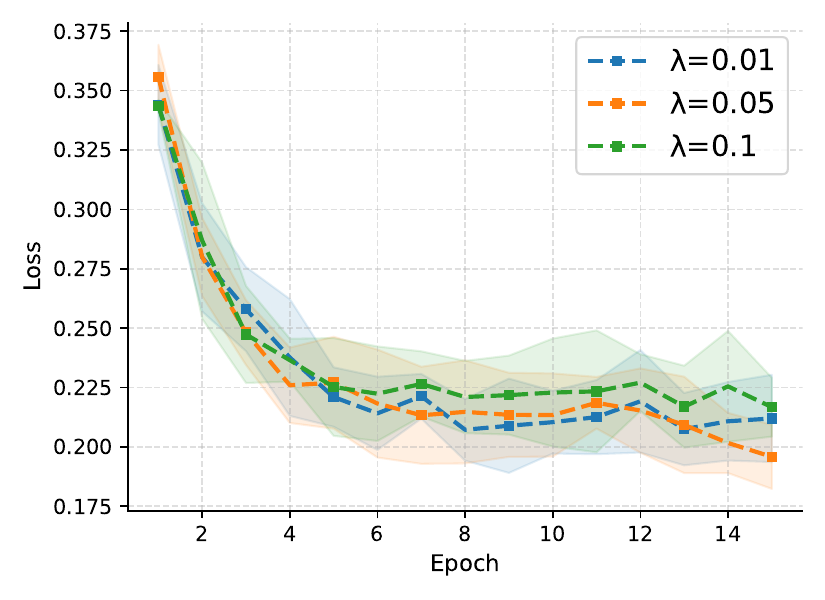}%
			\includegraphics[width=0.23\textwidth]{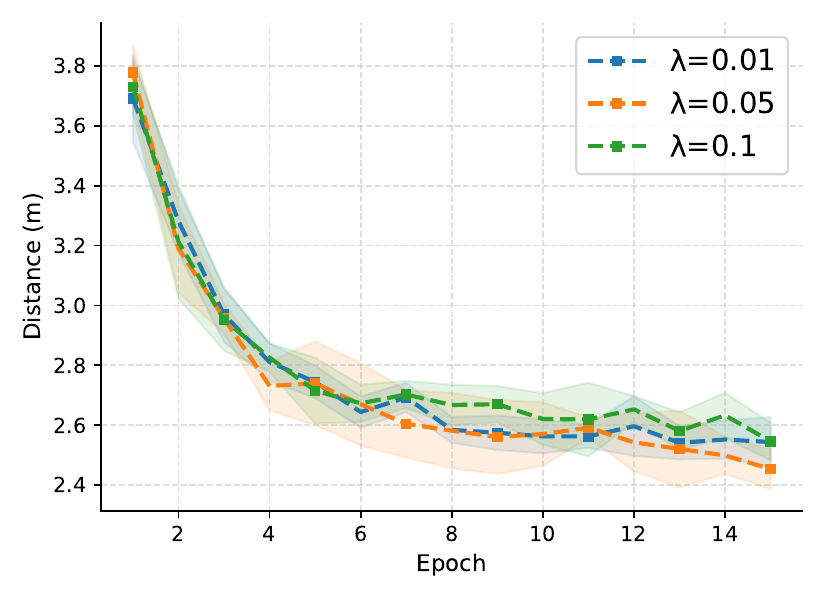}%
			\label{fig:Hyper_Tuneg4}%
		}
		\caption{Validation loss and distance for (a) learning rate ($\eta$), (b) batch size ($\mathcal{B}_S$), (c) state dimension ($d_{\text{state}}$), and (d) weight decay ($\lambda$).}
		\label{fig:Hyper_Tune}
	\end{figure*}
	\subsection{Fine-Tuning \& Analysis}

    \begin{figure}[t]
    \centering
    \includegraphics[width=\linewidth]{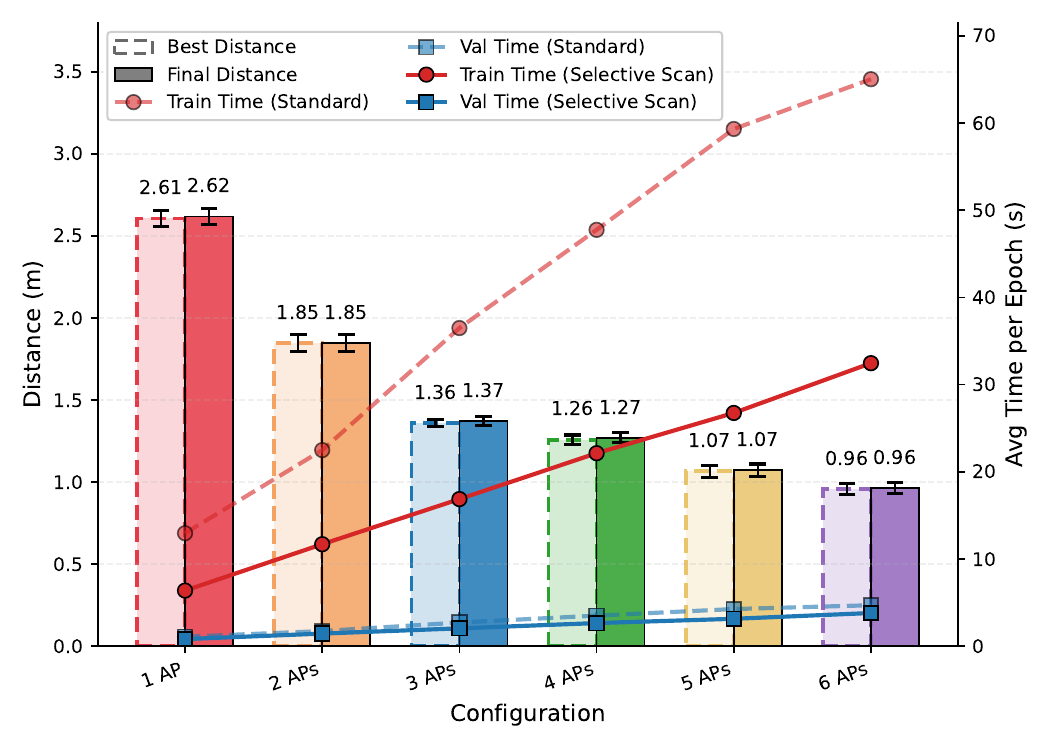}
    \caption{Obtained validation distance with different number of \glspl{AP}.}
    \label{fig:num_AP_sweep}
\end{figure}

	First, we fine-tune the hyper-parameters of the proposed LocAttMamba with a single \gls{AP} (i.e., a single Mamba block and no attention for fusion) for a faster tuning. We initialize the hyper-parameters based on \cite{F_GMamba}, and then tune several architectural and training hyper-parameters. Fig. \ref{fig:Hyper_Tune} shows the validation loss and validation distance across different learning rates $\eta = \left[0.0001, 0.0003, 0.001, 0.003, 0.01\right]$, batch sizes $\mathcal{B}_S = \left[8, 16, 32, 64\right]$, $d_{\text{state}} = \left[8, 16, 32\right]$, and regularization parameter $\lambda = \left[0.01, 0.05, 0.1\right]$. In particular, Fig. \ref{fig:Hyper_Tuneg1} illustrates that $\eta = 0.003$ and $\eta = 0.01$ obtain the lowest validation distance, while $\eta=0.003$ offers lower variance for more stable performance. Similarly, Fig. \ref{fig:Hyper_Tuneg2} shows the performance across various batch sizes, where both $\mathcal{B}_S=16$ and $\mathcal{B}_S=8$ show an almost comparable performance, with $\mathcal{B}_S =16$ having a slightly lower variance. Additionally, training with a $\mathcal{B}_S=8$ results in a longer training time without substantial improvement. Subsequently, we tune the dimension $d_{\text{state}}$, shown in Fig. \ref{fig:Hyper_Tuneg3}. Notably, to avoid prolonged tuning, we consider $d_{\text{inner}} = d_{\text{model}} = 2 \times d_{\text{state}}$. Fig. \ref{fig:Hyper_Tuneg3} demonstrates that $d_{\text{state}} = 16$ achieves the lowest validation loss and distance, while increasing the $d_{\text{state}}$ to $32$ not only increases the complexity of the model but also does not lead to any improvement. Finally, Fig. \ref{fig:Hyper_Tuneg4} shows the sweep over weight decay $\lambda$. It can be seen that this parameter has the lowest impact on the performance compared to previous one, with $\lambda=0.05$ obtaining a slightly lower error overall. The summary of the hyper-parameters are shown in Table \ref{tab:mamba_hyperparams}. These hyper-parameters are used symmetrically for all Mamba blocks. \par
	Next, we analyze the impact of adding more \glspl{AP} on the validation distance. Fig. \ref{fig:num_AP_sweep} shows the obtained validation distance with different number of \glspl{AP} in addition to the training and validation time. As it is expected, adding the information from more \glspl{AP} leads to performance enhancement. For instance, from 1 \gls{AP} to 2 \glspl{AP} we observe $29.12\%$ lower validation distance while increasing the training time by $94.99\%$ and validation time by $77.61\%$. Similarly, from 5 \glspl{AP} to 6 \glspl{AP}, the improvement is $10.28\%$ with training and validation time increasing by $17.59\%$ and $17.47\%$, respectively. Fig. \ref{fig:num_AP_sweep} also shows the impact of using selective scan \cite{gu2023mamba} instead of standard sequential computation of the states inside the Mamba blocks. For instance, using the selective scan can reduce the training and validation for 6 \glspl{AP} by $50.08\%$ and $18.97\%$, respectively.
	\subsection{Ablation Study}
	In this subsection, we conduct an ablation study-- using the validation set-- on fusion, model, training, and data features of the proposed model as:
	\begin{enumerate}
		\item \textbf{Average Fusion}: Using average instead of attention.
		\item \textbf{No Fusion}: 6 Mamba blocks are trained separately, and the final location is the average of their prediction. 
		\item \textbf{\gls{CIR}-Only}: Using only the \gls{CIR} as the input.
		\item \textbf{\gls{CIR} + \gls{TDoA}}: dropping the \gls{ToA}-offset.
		\item \textbf{Unidirectional Mamba}: Instead of having both Mamba and Reverse Mamba, we only keep the Mamba.
		\item \textbf{1D CNN/RNN}: Using \gls{1-D} \gls{CNN}/\gls{RNN} instead of the Mamba blocks. 
		\item \textbf{Warm-up Only}: Training only with the \gls{MSE} loss.
	\end{enumerate}
	
	Table \ref{tab:mamba_ablation} shows the performance of the above-mentioned ablations with the associated performance degradation. For instance, replacing the attention layer with the \emph{Average Fusion} leads to an almost $9.43\%$ degradation, while with \emph{No Fusion} the performance drops substantially by $51.76\%$. Similarly, removing \gls{TDoA} and \gls{ToA}-offset feature results in a $26.15\%$ performance loss, whereas only removing the \gls{ToA}-offset has a small $1.03\%$ degradation. Notably, since \gls{ToA}-offset is a single scalar feature, we keep that as the input feature since it will have minimum impact on the complexity. Subsequently, using only unidirectional Mamba or replacing the whole block with \gls{1-D} \gls{CNN} or \gls{RNN} can result in the performance degradation by $15.79\%$, $12.73\%$, and $12.73\%$, respectively. Finally, only using the \gls{MSE} loss shows a small degradation as low as $4.00\%$ compared to the full \gls{MSE} + \gls{NLL} training. 
		
	\subsection{Performance on The Test Set}
\begin{figure}[!t]
		\centering
		\includegraphics[width=\linewidth]{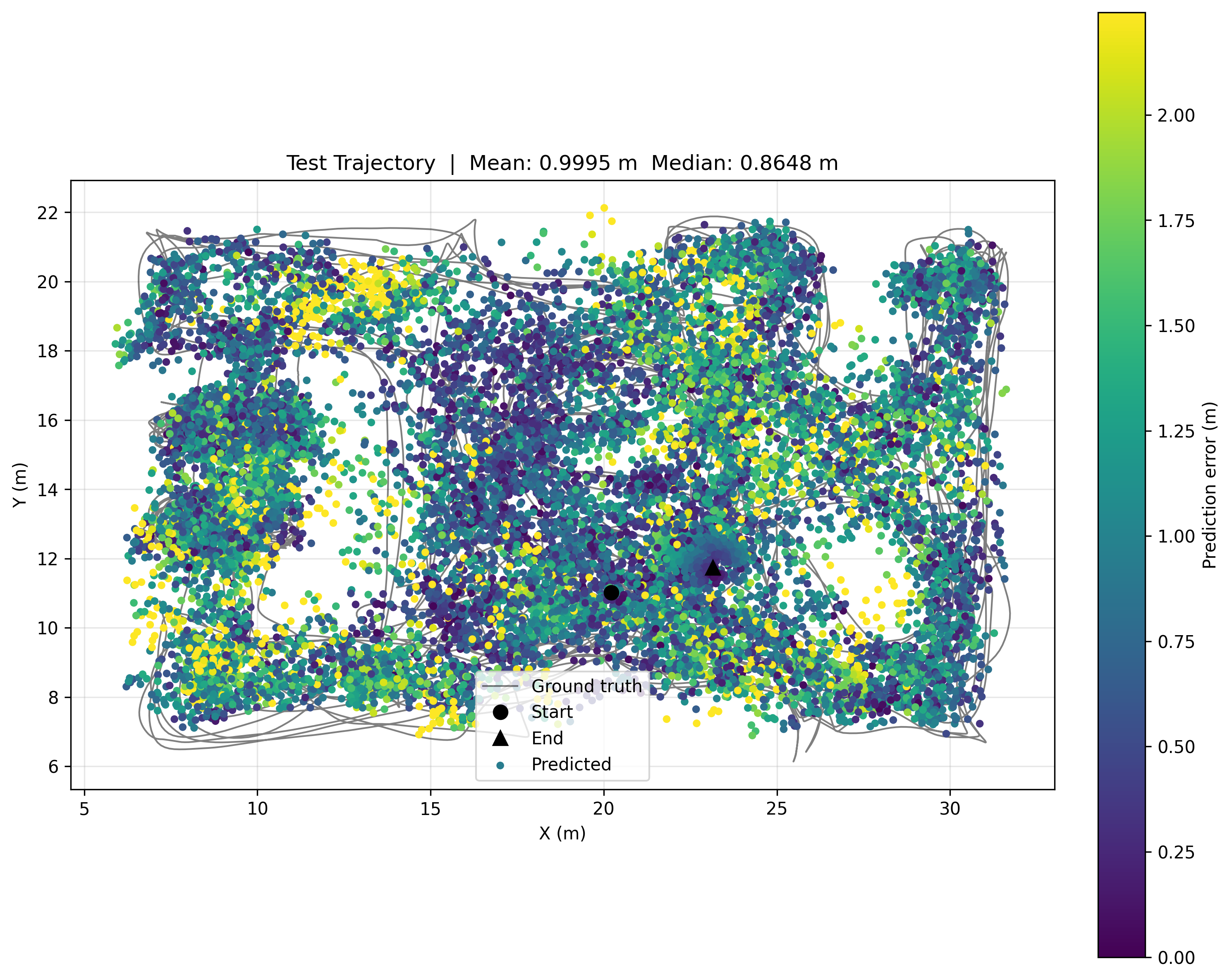}
		\caption{Error over the predicted locations on the test set.}
		\label{fig:test_error_trajectory}
	\end{figure}

	In this subsection, we evaluate the trained model on the unseen test set. Fig. \ref{fig:test_error_trajectory} illustrates the predicted locations against the ground-truth locations, where each point's color shows the prediction error. For instance, the points with darker colors have lower prediction errors, whereas the points with brighter colors close to yellow have higher prediction errors. The mean error on the test set obtained by the proposed LocAttMamba is $0.999$ m with median error of $0.865$ m\footnote{This number is obtained from a single run to visualize the predicted locations.}. \par
	\begin{table}[!t]
		\centering
		\caption{Ablation Study (6-\glspl{AP} Setting)}
		\label{tab:mamba_ablation}
		\resizebox{\linewidth}{!}{%
			\begin{tabular}{llcc}
				\hline
				\textbf{Ablation} & \textbf{Approach} & \textbf{Avg. Val. Dist.} & \textbf{Performance Degradation} \\
				\hline
				\multirow{2}{*}{Fusion}
				& Average Fusion   & $1.06 \pm 0.01$~m & 9.43\% \\
				& No Fusion   & $1.99 \pm 0.06$~m & 51.76\% \\
				\hline
				\multirow{2}{*}{Data} & CIR only & $1.30 \pm 0.04$~m & 26.15\% \\
				& CIR + TDoA & $0.97 \pm 0.04$~m & 1.03\% \\
				\hline
				\multirow{3}{*}{Model} & Unidirectional Mamba & $1.14 \pm 0.03$~m & 15.79\% \\
				& 1D CNN & $1.10 \pm 0.02$~m & 12.73\% \\
				& RNN & $1.10 \pm 0.03$~m & 12.73\% \\
				\hline
				\multirow{1}{*}{Training} & Warmup only & $1.00 \pm 0.01$~m & 4.00\% \\
				\hline
				\rowcolor[RGB]{245,193,0}
				\multicolumn{2}{l}{\textbf{LocAttMamba}} & $\mathbf{0.96} \boldsymbol{\pm} \mathbf{0.03}$\textbf{~m} & \textbf{---} \\
				\hline
			\end{tabular}%
		}
	\end{table}
	Furthermore, we evaluate the calibration of the predicted uncertainty using the normalized squared error in (\ref{eq:chi2}). Before calibration, the model is
	substantially overconfident, with $\chi^2 = 34.6$ and per-axis ratios of
	$13.43$ and $21.12$ along the $x$- and $y$-axes, respectively. Applying the
	per-axis temperature scaling of (\ref{eq:temp_scale}), with scales
	$s_x = 3.26$ and $s_y = 4.87$ fit on the validation set, reduces the
	test-set statistic to $\chi^2 = 2.15$, close to the ideal value of $2$. Here, the residual per-axis error-to-variance ratios are 1.11 and 0.75, indicating that the gross over-confidence is removed while the y-axis becomes mildly under-confident. The scales generalize well, where the validation-fitted values match the
	test-optimal scales within approximately $10\%$, confirming that the
	correction transfers to unseen data rather than overfitting the evaluation
	set. It is important to note that the post-hoc calibration does not impact the location head $\boldsymbol{\mu}$, and the temperature scaling is only applied to the uncertainty head.\par
\begin{figure}[!t]
		\centering
		\subfloat[$\bar{\boldsymbol{\alpha}}^{(0)}=0.2267$]{%
			\includegraphics[width=0.333\linewidth]{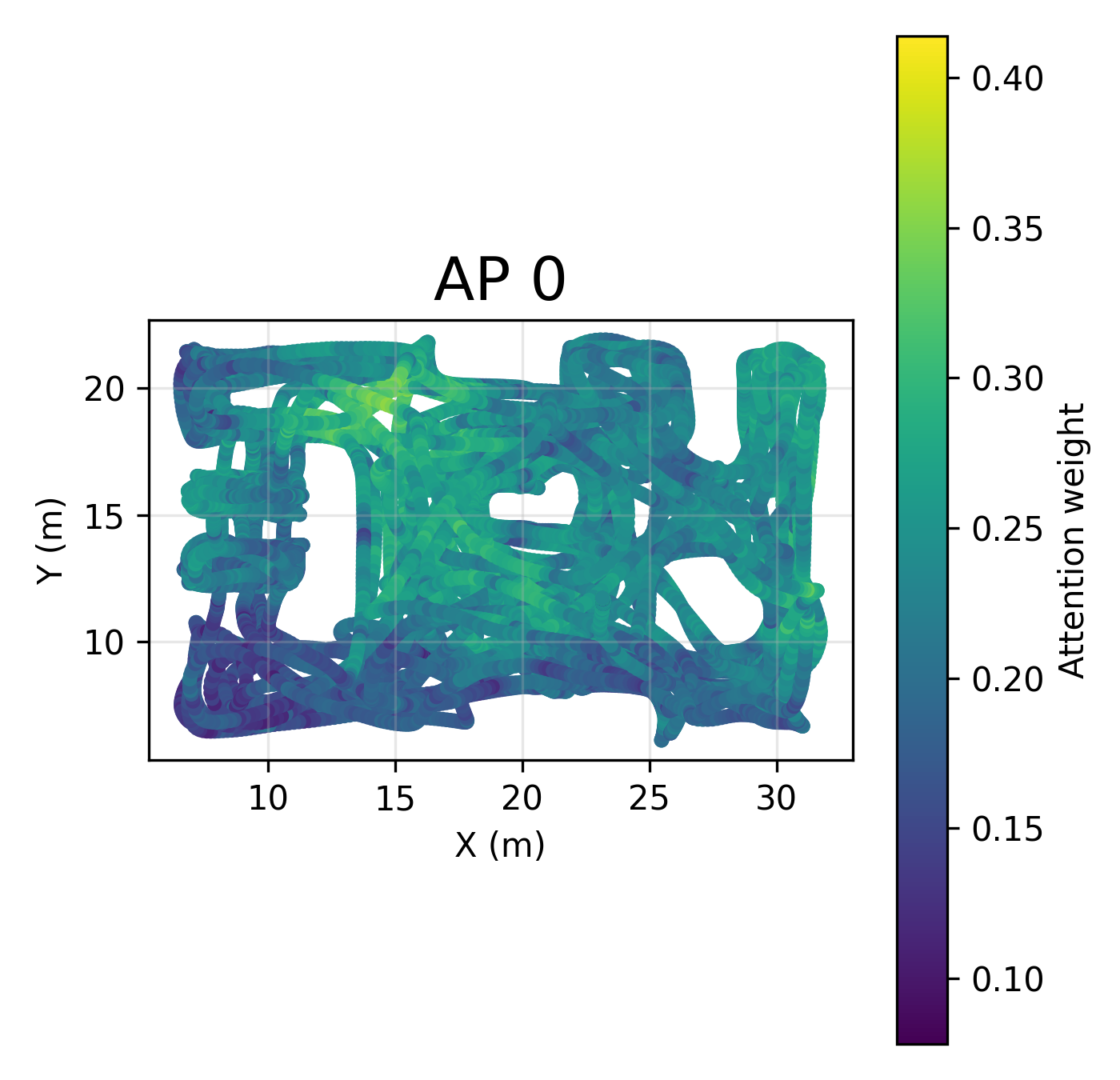}%
			\label{fig:attention_weight_AP0}%
		}%
		\subfloat[$\bar{\boldsymbol{\alpha}}^{(1)}=0.1408$]{%
			\includegraphics[width=0.333\linewidth]{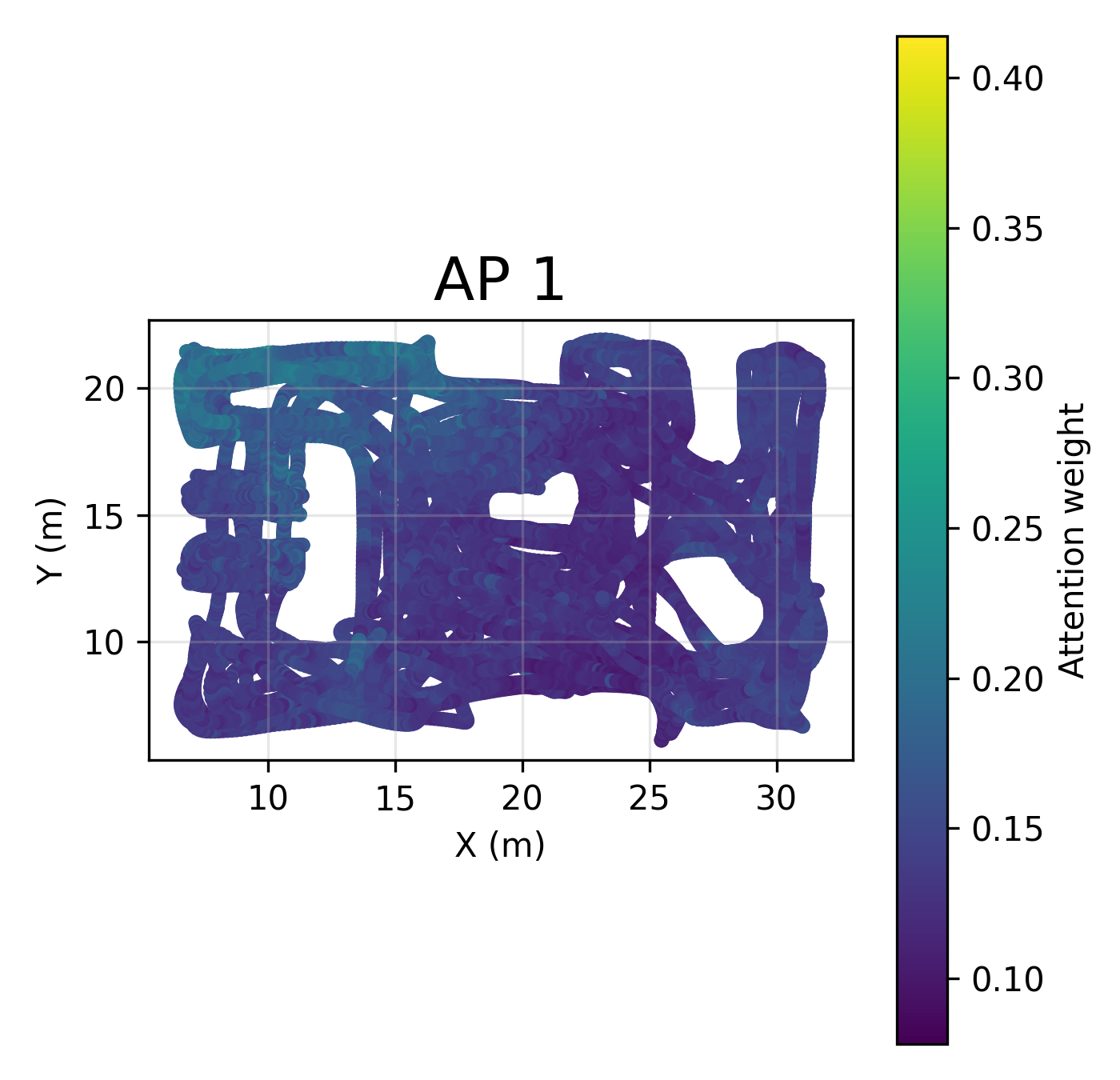}%
			\label{fig:attention_weight_AP1}%
		}%
		\subfloat[$\bar{\boldsymbol{\alpha}}^{(2)}=0.1476$]{%
			\includegraphics[width=0.333\linewidth]{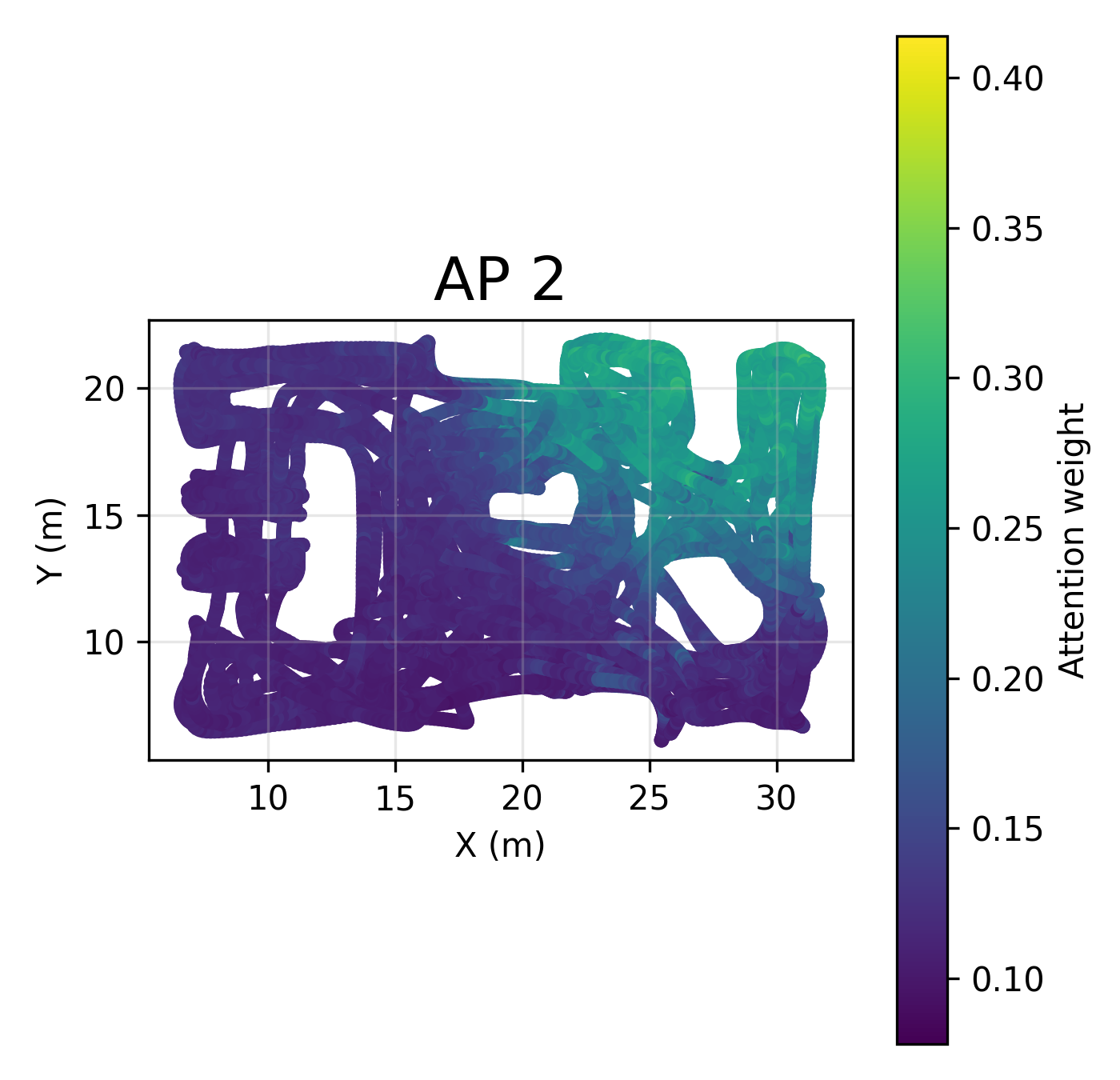}%
			\label{fig:attention_weight_AP3}%
		}
		
		
		\subfloat[$\bar{\boldsymbol{\alpha}}^{(3)}=0.1032$]{%
			\includegraphics[width=0.333\linewidth]{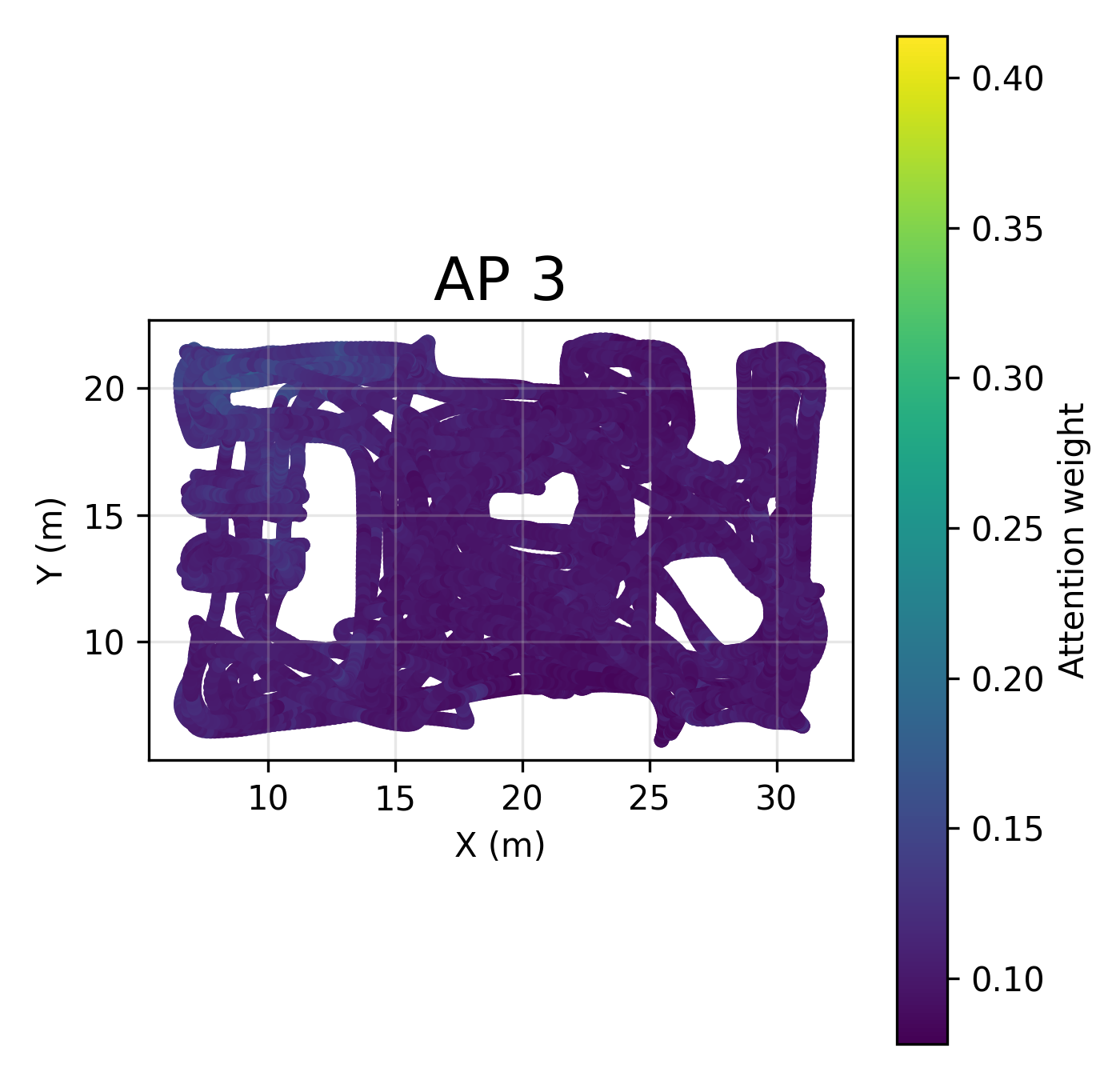}%
			\label{fig:attention_weight_AP4}%
		}%
		\subfloat[$\bar{\boldsymbol{\alpha}}^{(4)}=0.1784$]{%
			\includegraphics[width=0.333\linewidth]{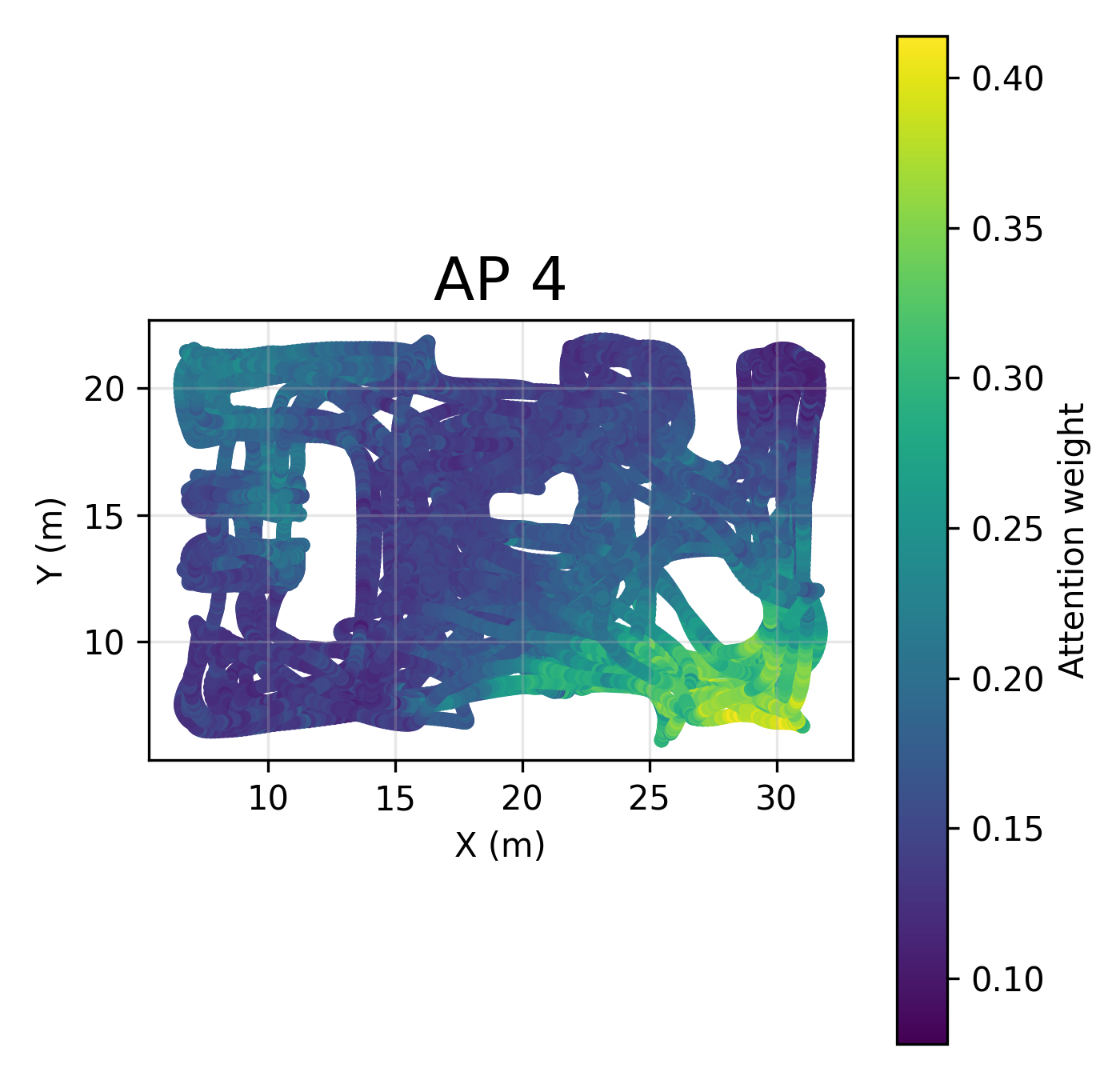}%
			\label{fig:attention_weight_AP5}%
		}%
		\subfloat[$\bar{\boldsymbol{\alpha}}^{(5)}=0.2032$]{%
			\includegraphics[width=0.333\linewidth]{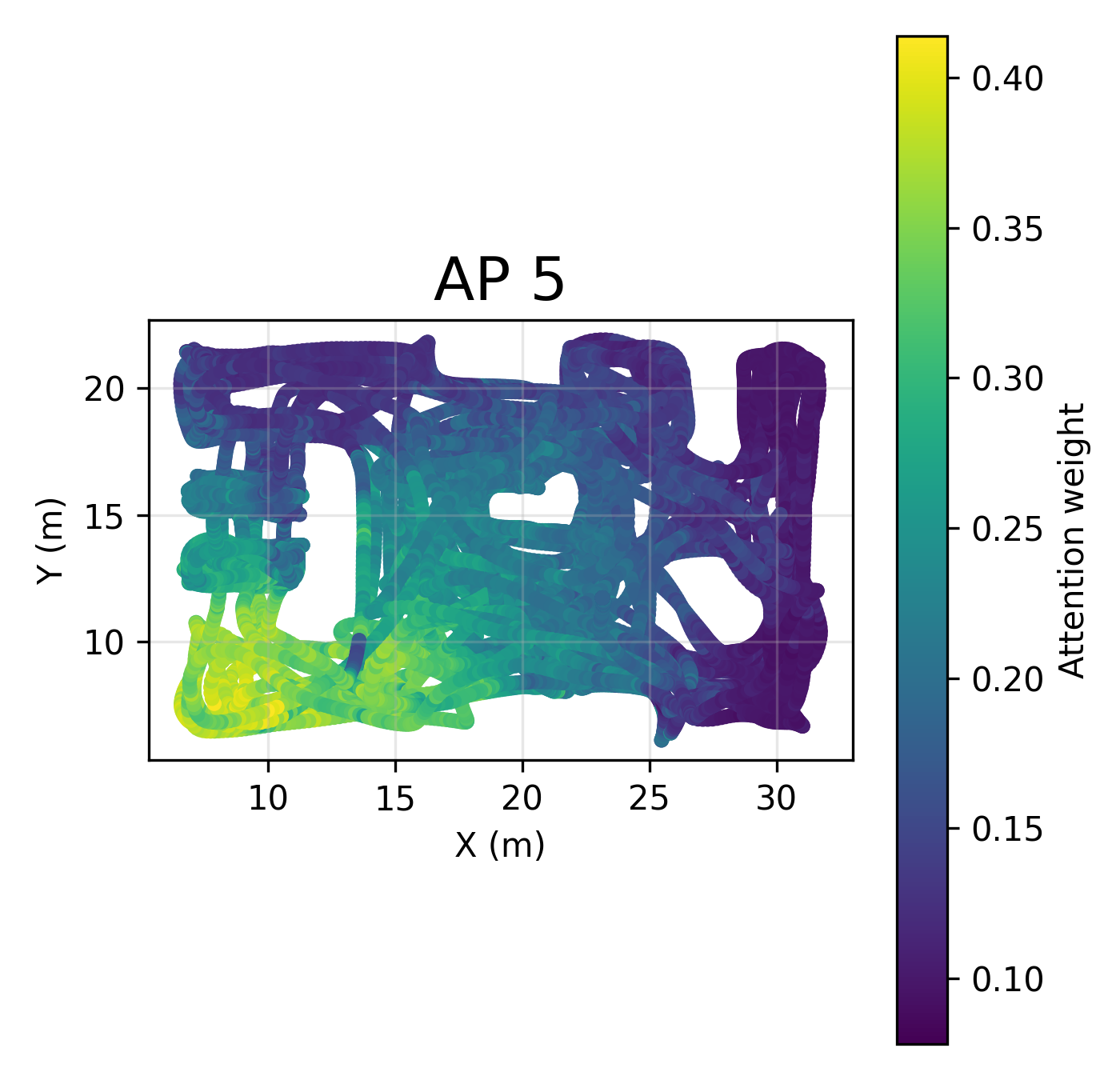}%
			\label{fig:attention_weight_AP6}%
		}
		
		\caption{Average attention weight for all \glspl{AP}.}
		\label{fig:average_attention_weight}
	\end{figure}

	To understand the contribution of each \gls{AP} in each region, the average attention weight $\bar{\boldsymbol{\alpha}}$ in (\ref{eq:attention_weight}) over all test time steps is illustrated in Fig. \ref{fig:average_attention_weight}. It can be seen that \glspl{AP} have different levels of impact in different regions, based on their proximity and line-of-sight availability. For instance, \gls{AP} number 5 has a higher attention weight in the lower left regions (shown in Fig. \ref{fig:attention_weight_AP6}), whereas \gls{AP} 3 has the lowest average attention weight, due to the presence of an obstacle in front of the \gls{AP} (shown in Fig. \ref{fig:attention_weight_AP4}). In Section \ref{subsec:noise}, we will discuss the impact of noise on the attention weights to better observe the weighting mechanism in the fusion.
	
	\subsection{Benchmark Comparison}
	After evaluating the performance of the proposed LocAttMamba, we compare the performance against several existing benchmarks as:
	\begin{enumerate}
		\item \textbf{\gls{SVM}}: A conventional \gls{ML} approach.
		\item \textbf{Bi-LSTM \cite{bi-lstm}}: a Bi-LSTM-based approach for indoor fingerprinting.
		\item \textbf{\gls{ACNN} \cite{ACCN}}: An Autoencoder-based \gls{CNN} indoor localization.
		\item \textbf{L-SwiGLU-T \cite{L-SwiGLU-T}}: A lightweight transformer-based approach for indoor localization.
		\item \textbf{Swin-Loc \cite{SwinLoc}}: A heavy-weight transformer architecture for sensing and localization.
		\item \textbf{Sec5GLoc \cite{Sec5GLoc}}: A \gls{CNN}-attention approach for indoor localization.
		\item \textbf{MambaLoc \cite{MambaLoc}}: A Mamba-based indoor localization, trained as a standalone model since the considered datasets do not include the paired modality used for distillation in \cite{MambaLoc}.
	\end{enumerate}
	\par 
	For a more reliable output, we also consider the \gls{UWB} localization dataset in \cite{fraunhofer_dataset}, containing \gls{TDoA} and \gls{ToA}-offset, and \gls{CIR} with $\nu_{\text{\gls{UWB}}}=366$. This dataset also includes $N_{\text{\gls{AP}, \gls{UWB}}} = 6$, where the data is collected with a bandwidth of $499.2$ MHz and a carrier frequency of $4$ GHz with $T_{\text{train,\gls{UWB}}} = 18762$ and $T_{\text{test,\gls{UWB}}} = 3520$. \par
	\begin{table*}[!t]
		\centering
		\caption{Benchmark Comparison across 5G and UWB scenarios.}
		\label{tab:benchmark test}
		\resizebox{\textwidth}{!}{%
			\begin{tabular}{llccccccc}
				\hline\hline\noalign{\smallskip}
				\textbf{Dataset} & \textbf{Model} & \textbf{Mean} & \textbf{Median} & \textbf{P75} & \textbf{P90} & \textbf{Inf. Time (ms/sample)} & \textbf{Multiply-Acc. (M/sample)} & \textbf{\# Params} \\
				\noalign{\smallskip}\hline\noalign{\smallskip}
				
				\multirow{8}{*}{\textbf{5G}} 
				& SVM & $3.166 \pm 0.000$ & $2.844 \pm 0.000$& $4.144 \pm 0.000$ & $5.700 \pm 0.000$ & $51.051 \pm 0.870$ & $43.936$ & N/A \\
				& Bi-LSTM \cite{bi-lstm}& $1.530 \pm 0.046$ & $1.274 \pm 0.042$ & $1.995 \pm 0.069$ & $2.917 \pm 0.106$ & $0.065 \pm 0.001$ & $2.151$ & $2,146,690$ \\
				& \gls{ACNN} \cite{ACCN}& $ 2.153 \pm 0.055$ & $1.764 \pm 0.040$ & $2.646 \pm 0.050$ & $3.755 \pm 0.090$ & $1.198 \pm 0.030$ & $13.592$ & $13,880,846$ \\
				& L-SwiGLU-T \cite{L-SwiGLU-T}& $1.467 \pm 0.036$ & $1.118 \pm 0.031$ & $1.811 \pm 0.041$ & $2.858 \pm 0.066$ & $5.325 \pm 0.037$ & $1.288$ & $218,310$ \\
				& Swin-Loc \cite{SwinLoc}& $1.113 \pm 0.012$ & $0.976 \pm 0.026$ & $1.467 \pm 0.042$ & $1.955 \pm 0.052$ & $18.895 \pm 0.102$ & $905.770$ & $4,005,855$ \\
				& Sec5GLoc \cite{Sec5GLoc}& $1.122 \pm 0.019$ & $\mathbf{0.825} \boldsymbol{\pm} \mathbf{0.017}$ & $1.345 \pm 0.027$ & $2.132 \pm 0.023$ & $1.863 \pm 0.026$ & $27.945$ & $211,986$ \\
				& MambaLoc \cite{MambaLoc}& $2.012 \pm 0.042$ & $1.732 \pm 0.028$ & $2.661 \pm 0.014$ & $3.751 \pm 0.018$ & $0.261 \pm 0.004$ & $0.721$ & $657,328$ \\
				\rowcolor[RGB]{245,193,0}
				& \textbf{LocAttMamba} & $\mathbf{1.004} \boldsymbol{\pm} \mathbf{0.010}$ & $0.874 \pm 0.014$& $\mathbf{1.321} \boldsymbol{\pm} \mathbf{0.018}$ & $\mathbf{1.840} \boldsymbol{\pm} \mathbf{0.015}$ & $1.171 \pm 0.027$ & $0.446$ & $344,804$ \\
				
				\noalign{\smallskip}\hline\noalign{\smallskip}
				
				\multirow{8}{*}{\textbf{UWB}} 
				& SVM & $1.595 \pm 0.000$ & $1.393 \pm 0.000$ & $2.022 \pm 0.000$ & $2.735 \pm 0.000$ & $136.706 \pm 5.209$ & $123.206$ & N/A \\
				& Bi-LSTM \cite{bi-lstm} & $1.656 \pm 0.026$ & $1.585 \pm 0.028$ & $2.180 \pm 0.039$ & $2.811 \pm 0.061$ & $0.069 \pm 0.001$ & $5.075$ & $5,071,234$ \\
				& \gls{ACNN} \cite{ACCN} & $0.850 \pm 0.009$ & $0.762 \pm 0.019$ & $1.111 \pm 0.015$ & $1.531 \pm 0.012$ & $1.214 \pm 0.038$ & $19.441$ & $25,581,878$ \\
				& L-SwiGLU-T \cite{L-SwiGLU-T}& $0.636 \pm 0.021$ & $\mathbf{0.486} \boldsymbol{\pm} \mathbf{0.004}$ & $\mathbf{0.759} \boldsymbol{\pm}\mathbf{0.017}$ & $1.135 \pm 0.035$ & $5.243 \pm 0.053$ & $1.390$ & $235,446$ \\
				& Swin-Loc \cite{SwinLoc}& $1.632 \pm 0.032$ & $1.519 \pm 0.022$ & $2.166 \pm 0.065$ & $2.773 \pm 0.052$ & $18.952 \pm 0.362$ & $2,604.063$ & $9,895,487$ \\
				& Sec5GLoc \cite{Sec5GLoc}& $0.781 \pm 0.048$ & $0.537 \pm 0.047$ & $0.868 \pm 0.062$ & $1.409 \pm 0.078$ & $2.034 \pm 0.048$ & $78.485$ & $211,986$ \\
				& MambaLoc \cite{MambaLoc}& $1.953 \pm 0.057$ & $1.816 \pm 0.055$ & $2.592 \pm 0.080$ & $3.374 \pm 0.108$ & $0.253 \pm 0.005$ & $1.087$ & $1,022,896$ \\
				\rowcolor[RGB]{245,193,0}
				& \textbf{LocAttMamba} & $\mathbf{0.599} \boldsymbol{\pm} \mathbf{0.011}$ & $0.560 \pm 0.007$ & $0.792 \pm 0.017$ & $\mathbf{1.026} \boldsymbol{\pm} \mathbf{0.029}$ & $1.191 \pm 0.014$ & $0.953$ & $436,196$ \\
				
				\noalign{\smallskip}\hline\hline
			\end{tabular}%
		}
	\end{table*}

	To compare the models, we consider mean, median, \gls{P75}, and \gls{P90} error. Additionally, the complexity of each model is evaluated through its inference time and multiply-accumulate operations. All the models are given the same input features as \gls{CIR}, \gls{TDoA}, and \gls{ToA}-offset. \par 
	Table \ref{tab:benchmark test} shows the overall performance of the benchmarks on both \gls{5G} and \gls{UWB} test sets, averaged over 4 independent runs. It can be seen that the proposed LocAttMamba obtains the best mean error on both datasets, where it outperforms the second best model by $9.79\%$ (Swin-Loc \cite{SwinLoc}) in \gls{5G} and $5.82\%$ (L-SwiGLU-T \cite{L-SwiGLU-T}) in \gls{UWB}. On the other hand, the Sec5GLoc \cite{Sec5GLoc} obtains a better median performance on \gls{5G} dataset by $5.61\%$, while having higher \gls{P75} and \gls{P90} error by $1.82\%$ and $13.70\%$, respectively. Similarly, L-SwiGLU-T \cite{L-SwiGLU-T} has a lower median and \gls{P75} error by $13.27\%$ and $4.17\%$, its \gls{P90} error is higher by $9.60\%$. \par
	In terms of the complexity of benchmarks, the proposed LocAttMamba has the lowest multiply-accumulate operations compared to all other benchmarks, while having third lowest inference time among the benchmarks. In particular, the Bi-LSTM \cite{bi-lstm} benchmark has much lower inference time by almost $18$ times compared to the proposed LocAttMamba, however, the performance of this benchmark is much worse by almost $34.40\%$ on \gls{5G} dataset and $2.76$ times on the \gls{UWB} dataset, completely outweighing its low inference time. Compared to Sec5GLoc \cite{Sec5GLoc}, which has a $5.61\%$ lower median error on the \gls{5G} dataset, the proposed LocAttMamba has $37.10\%$ lower inference time and almost $63$ times lower multiply-accumulate operations on the \gls{5G} dataset. Similarly, while L-SwiGLU-T \cite{L-SwiGLU-T} has lower median and \gls{P75} error in \gls{UWB} dataset, it has roughly 1.5 times more multiply-accumulate operations and almost 4.4 times higher inference time, making the model inappropriate for real-time deployment. It is important to note that all the models are tested using T4 GPU and batch size = 1 for inference time.\par
\begin{figure}[t]
    \centering
    \includegraphics[width=\linewidth]{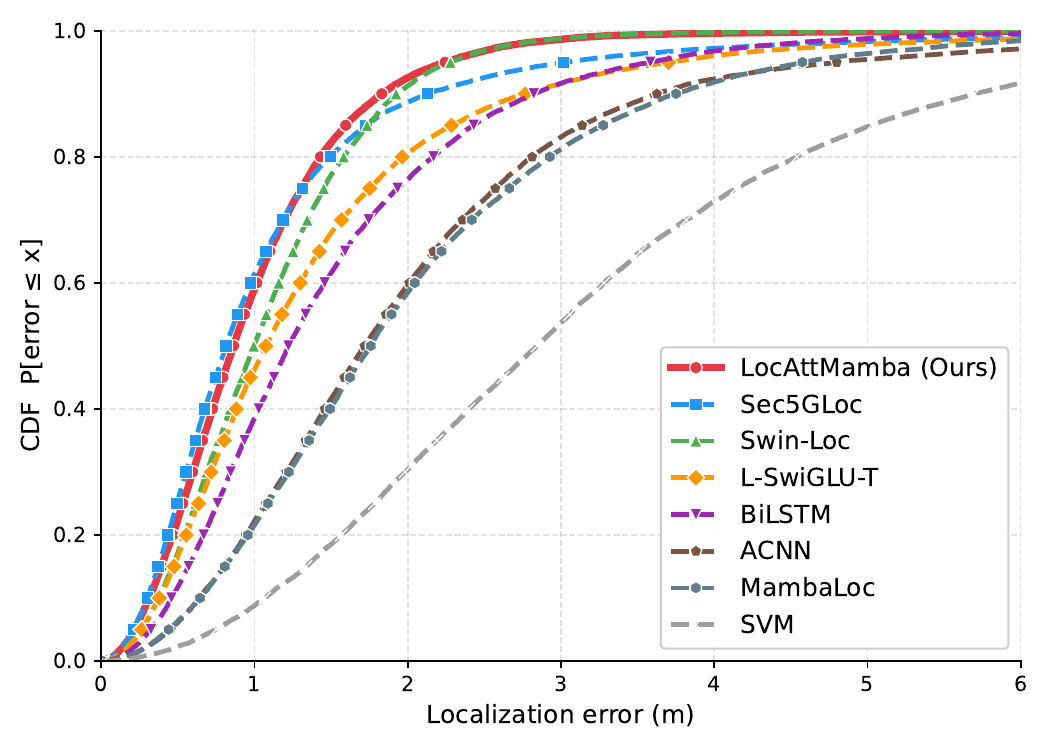}
    \caption{Error \gls{CDF} on \gls{5G} test set.}
    \label{fig:5G_analysis}
\end{figure}

	To better illustrate the performance, the \gls{CDF} of error of the benchmarks on the \gls{5G} test set is shown in Fig. \ref{fig:5G_analysis}. It can be seen that this figure confirms the results in Table \ref{tab:benchmark test}, where for instance, the Sec5GLoc \cite{Sec5GLoc} has a lower median with its \gls{CDF} more leaned towards zero, while the proposed LocAttMamba approach has better \gls{P75} and \gls{P90} errors. Additionally, it can be seen that the SVM baseline has the worst \gls{CDF} curve compared to other approaches.
	\subsection{Noise Robustness}
	\label{subsec:noise}
	While the proposed LocAttMamba is superior on nearly all complexity and error metrics, it is important to consider real-world scenarios where the input \gls{CIR} may face \gls{SNR} degradation. Thus, in this subsection, we evaluate the robustness of the trained models under noisy \gls{CIR} inputs. The \gls{CIR} power at time step $t$, for \gls{AP} $i$ is written as: 
	\begin{equation}
		P_i^{(t)} = \frac{1}{\nu}\sum_{b=1}^{\nu}\big|h_{i,b}^{(t)}\big|^2 .
	\end{equation}

	Subsequently, noise power per complex tap for a target \gls{SNR} (in dB), and per-component standard deviation:
	\begin{equation}
		N_i^{(t)} = P_i^{(t)} \, 10^{-\mathrm{SNR}^{\mathrm{dB}}/10},
		\qquad
		\sigma_i^{(t)} = \sqrt{N_i^{(t)}/2}.
	\end{equation}
	Then, complex additive white Gaussian noise is added as:
	\begin{equation}
		\begin{split}
			n_{i,b}^{(t)} = \Re\left(n_{i,b}^{(t)}\right) + j\,\Im\left(n_{i,b}^{(t)}\right),
			\quad 
			n_{i,b}^{(t)\Re}, n_{i,b}^{(t)\Im} \stackrel{\text{i.i.d.}}{\sim} \mathcal{N}\!\big(0,\sigma_i^{(t)^2}\big),
		\end{split}
	\end{equation}
	
	Degraded \gls{CIR} $\tilde{h}_{i, b}^{(t)} \in \mathbb{C}$ can be written as:
	\begin{equation}
		\begin{split}
			\tilde{h}_{i,b}^{(t)} = h_{i,b}^{(t)} + n_{i,b}^{(t)}
			\;\Longleftrightarrow\; &
			\Re\left(\tilde{h}_{i,b}^{(t)}\right) = \Re\left(h_{i,b}^{(t)}\right) + \Re\left(n_{i,b}^{(t)}\right), \\
			&
			\Im\left(\tilde{h}_{i,b}^{(t)}\right) = \Im\left(h_{i,b}^{(t)}\right) + \Im\left(n_{i,b}^{(t)}\right).
		\end{split}
	\end{equation}


\begin{figure}[!t]
	\centering
	\subfloat[\gls{5G}]{%
		\includegraphics[width=0.95\linewidth]{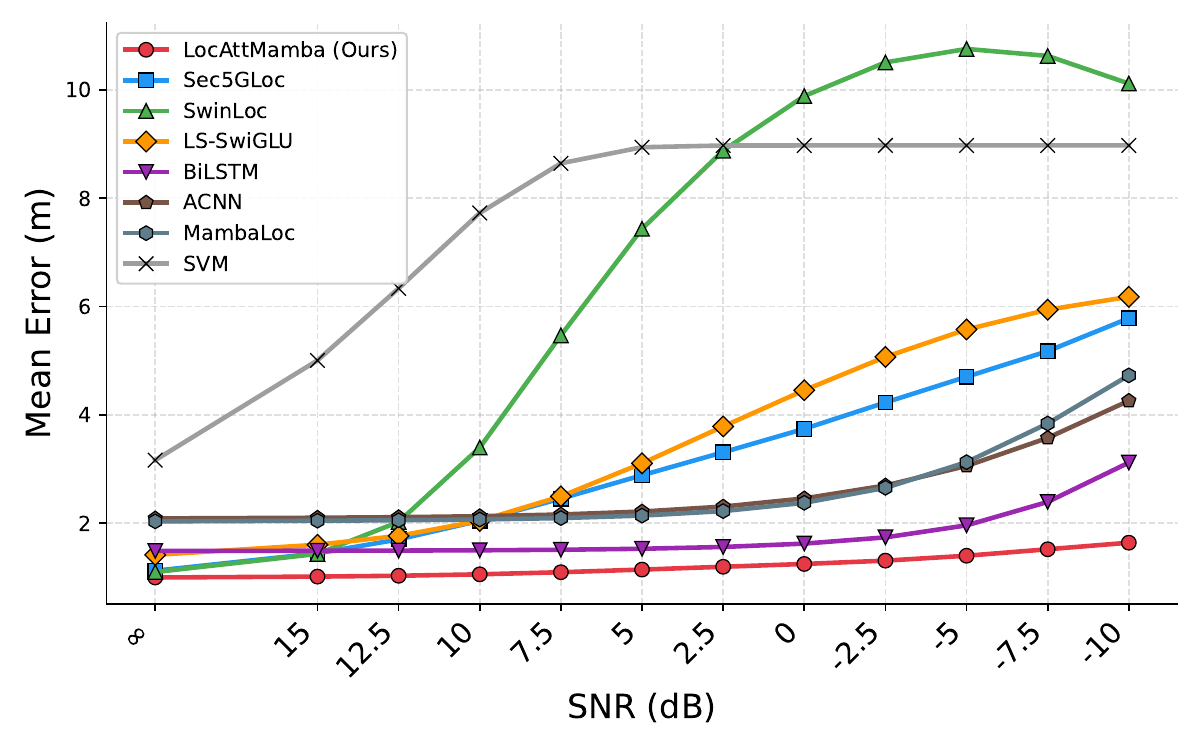}%
		\label{fig:error_SNR_5G}%
	}
	\\ 
	\subfloat[\gls{UWB}]{%
		\includegraphics[width=0.95\linewidth]{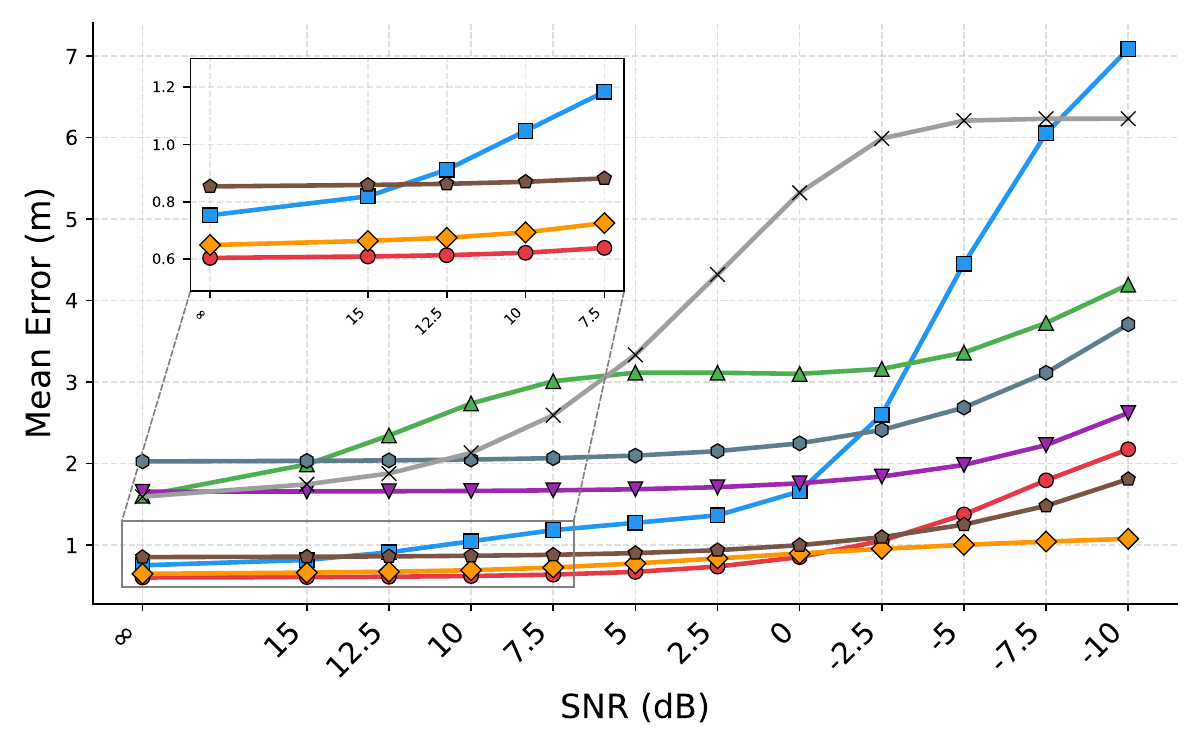}%
		\label{fig:error_SNR_UWB}%
	}
	\caption{Mean error versus degraded \gls{SNR} in (a) \gls{5G} and (b) \gls{UWB} datasets.}
	\label{fig:single_AP_SNR}
\end{figure}

		\begin{table*}[!t]
		\centering
		\caption{Robustness Evaluation: Absolute Mean Error (m) at a fixed degraded \gls{SNR} = 5 dB under varying \gls{AP} \gls{CIR} distortions}
		\label{tab:robustness_eval}
		\resizebox{\textwidth}{!}{%
			\begin{tabular}{l cccc c cccc}
				\hline\hline\noalign{\smallskip}
				\multirow{2}{*}{\textbf{Model}} & \multicolumn{4}{c}{\textbf{5G}} & \hspace{2.5em} & \multicolumn{4}{c}{\textbf{UWB}} \\
				\cline{2-5} \cline{7-10} \noalign{\smallskip}
				& \textbf{Clean} & \textbf{1 \gls{AP}} & \textbf{3 \glspl{AP}} & \textbf{5 \glspl{AP}} & & \textbf{Clean} & \textbf{1 \gls{AP}} & \textbf{3 \glspl{AP}} & \textbf{5 \glspl{AP}} \\
				\noalign{\smallskip}\hline\noalign{\smallskip}
				
				SVM & $3.166$ & $8.941 \pm 0.000$ & $8.974 \pm 0.000$ & $8.974 \pm 0.000$ & & $1.595$ & $3.335 \pm 0.005$ & $5.501 \pm 0.003$ & $6.020 \pm 0.003$ \\[1ex]
				Bi-LSTM \cite{bi-lstm} & $1.530$ & $1.570 \pm 0.006$ & $1.603 \pm 0.014$ & $\mathbf{1.641 \pm 0.010}$ & & $1.656$ & $1.686 \pm 0.010$ & $1.783 \pm 0.012$ & $1.873 \pm 0.026$ \\[1ex]
				\gls{ACNN} \cite{ACCN} & $2.153$ & $2.217 \pm 0.005$ & $2.395 \pm 0.007$ & $2.483 \pm 0.007$ & & $0.850$ & $0.902 \pm 0.003$ & $1.016 \pm 0.005$ & $1.097 \pm 0.010$ \\[1ex]
				L-SwiGLU-T \cite{L-SwiGLU-T} & $1.467$ & $3.107 \pm 0.002$ & $4.491 \pm 0.015$ & $5.906 \pm 0.013$ & & $0.636$ & $0.775 \pm 0.006$ & $1.329 \pm 0.016$ & $2.686 \pm 0.012$ \\[1ex]
				Swin-Loc \cite{SwinLoc} & $1.113$ & $7.436 \pm 0.028$ & $17.635 \pm 0.009$ & $13.717 \pm 0.012$ & & $1.632$ & $3.117 \pm 0.013$ & $3.818 \pm 0.011$ & $4.734 \pm 0.012$ \\[1ex]
				Sec5GLoc \cite{Sec5GLoc}& $1.122$ & $2.886 \pm 0.001$ & $5.747 \pm 0.002$ & $9.363 \pm 0.001$ & & $0.781$ & $1.274 \pm 0.004$ & $3.365 \pm 0.007$ & $6.990 \pm 0.003$ \\[1ex]
				MambaLoc \cite{MambaLoc} & $2.012$ & $2.142 \pm 0.002$ & $2.330 \pm 0.006$ & $2.456 \pm 0.010$ & & $1.953$ & $2.099 \pm 0.010$ & $2.343 \pm 0.020$ & $2.477 \pm 0.015$ \\[1ex]
				
				\hline\noalign{\smallskip}
				\rowcolor[RGB]{245,193,0}
				\textbf{LocAttMamba} & $\mathbf{1.004}$ & $\mathbf{1.146 \pm 0.004}$ & $\mathbf{1.448 \pm 0.006}$ & $1.736 \pm 0.014$ & & $\mathbf{0.599}$ & $\mathbf{0.673 \pm 0.012}$ & $\mathbf{0.762 \pm 0.015}$ & $\mathbf{0.849 \pm 0.021}$ \\
				
				\noalign{\smallskip}\hline\hline
			\end{tabular}%
		}
	\end{table*}
	We consider multiple-\gls{AP} $=\left\{1, 3, 5\right\}$ \gls{CIR} degradation over a single \gls{SNR} = 5 dB, and single-\gls{AP} \gls{CIR} degradation over range of \glspl{SNR}. \par
	Table \ref{tab:robustness_eval} shows the performance of the benchmarks under varying \gls{AP} distortions. For more reliable results, we consider adding 4 different noise seeds and average the results. It can be seen that the proposed LocAttMamba approach maintains its superiority on both datasets under multiple-\gls{AP} noise degradation. For instance, in \gls{5G} dataset and 3-\gls{AP} distortion, the proposed LocAttMamba obtains $37.85\%$, $74.80\%$, $67.76\%$, and $9.64\%$ lower mean error compared to MambaLoc \cite{MambaLoc}, Sec5GLoc \cite{Sec5GLoc}, L-SwiGLU-T \cite{L-SwiGLU-T}, and Bi-LSTM \cite{bi-lstm}, respectively. Similarly, on the \gls{UWB} dataset and 5-\gls{AP} \gls{CIR} degradation, the proposed LocAttMamba achieves $82.05\%$, $54.66\%$, and $22.62\%$ lower mean error than Swin-Loc \cite{SwinLoc}, Bi-LSTM \cite{bi-lstm}, and \gls{ACNN} \cite{ACCN}, respectively. Notably, it can be observed that the Bi-LSTM \cite{bi-lstm} approach has a lower mean error in \gls{5G} dataset and 5-\gls{AP} distortion by $5.48\%$, while having an inferior performance in all other settings. \par

	Subsequently, we consider degrading a single \gls{AP}\footnote{The same \gls{AP} is degraded for all algorithms.} and sweep over different \gls{SNR} values = $\left[15: 2.5: -10\right]$. Fig. \ref{fig:single_AP_SNR} shows the mean error versus obtained \gls{SNR} for both \gls{5G} and \gls{UWB} datasets. Particularly, the proposed LocAttMamba maintains its performance in almost all \glspl{SNR} across both datasets. For instance, in a moderate noise of \gls{SNR} = 7.5 dB, the proposed LocAttMamba has $27.43\%$ lower error than Bi-LSTM \cite{bi-lstm} in \gls{5G} and $11.96\%$ lower error than L-SwiGLU-T \cite{L-SwiGLU-T} in \gls{UWB} dataset as second best algorithms. Similarly, in an extreme noise of \gls{SNR} = -10 dB, LocAttMamba obtains a much lower error than L-SwiGLU-T \cite{L-SwiGLU-T} and Bi-LSTM \cite{bi-lstm} by $73.44\%$ and $47.38\%$, respectively, in the \gls{5G} dataset. However, in the \gls{UWB} dataset, L-SwiGLU-T \cite{L-SwiGLU-T} and \gls{ACNN} \cite{ACCN} achieve a lower error after \gls{SNR} = -2.5 dB than LocAttMamba, where at \gls{SNR} = -10 dB, they have $50.50\%$ and $16.85\%$ lower error, respectively, compared to LocAttMamba. \par

\begin{figure}[!t]
		\centering
		\subfloat[$\bar{\boldsymbol{\alpha}}^{(0)}=0.2267$]{%
			\includegraphics[width=0.32\linewidth]{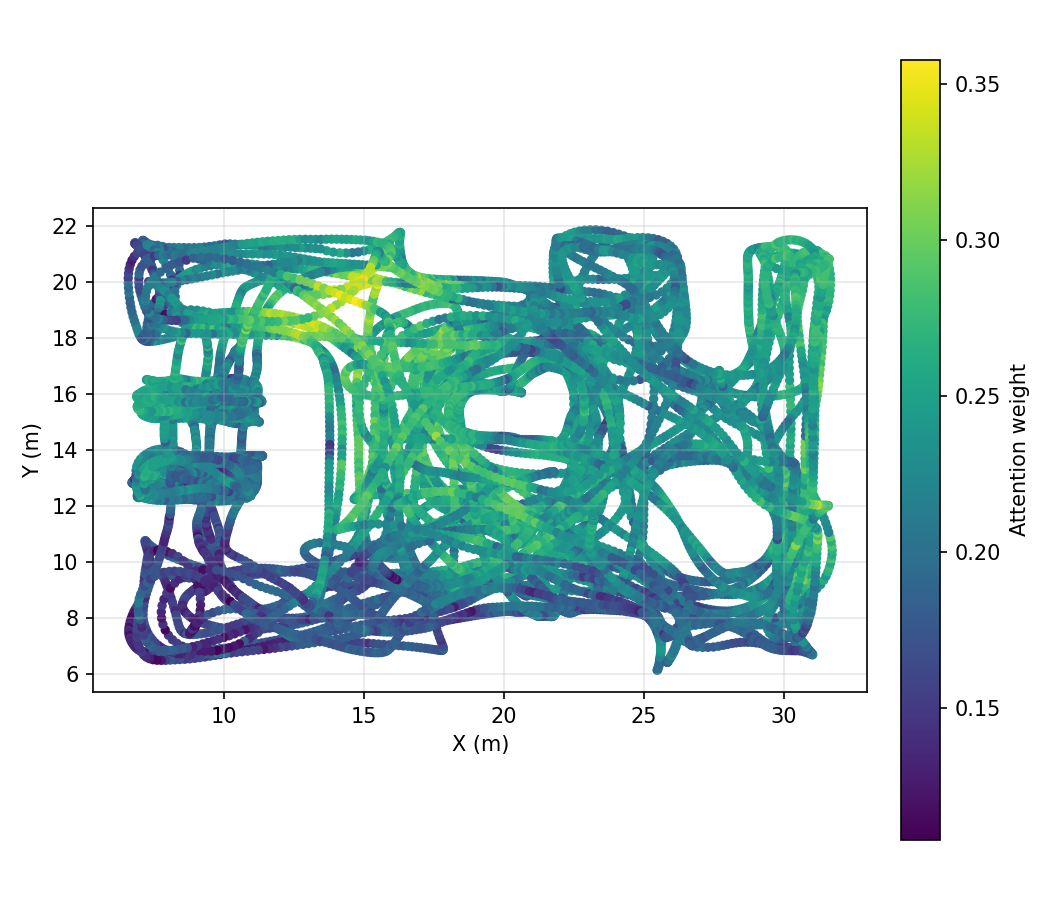}%
			\label{fig:noise_attention_weight_c}%
		}%
		\subfloat[$\bar{\boldsymbol{\alpha}}^{(0)}=0.2057$]{%
			\includegraphics[width=0.32\linewidth]{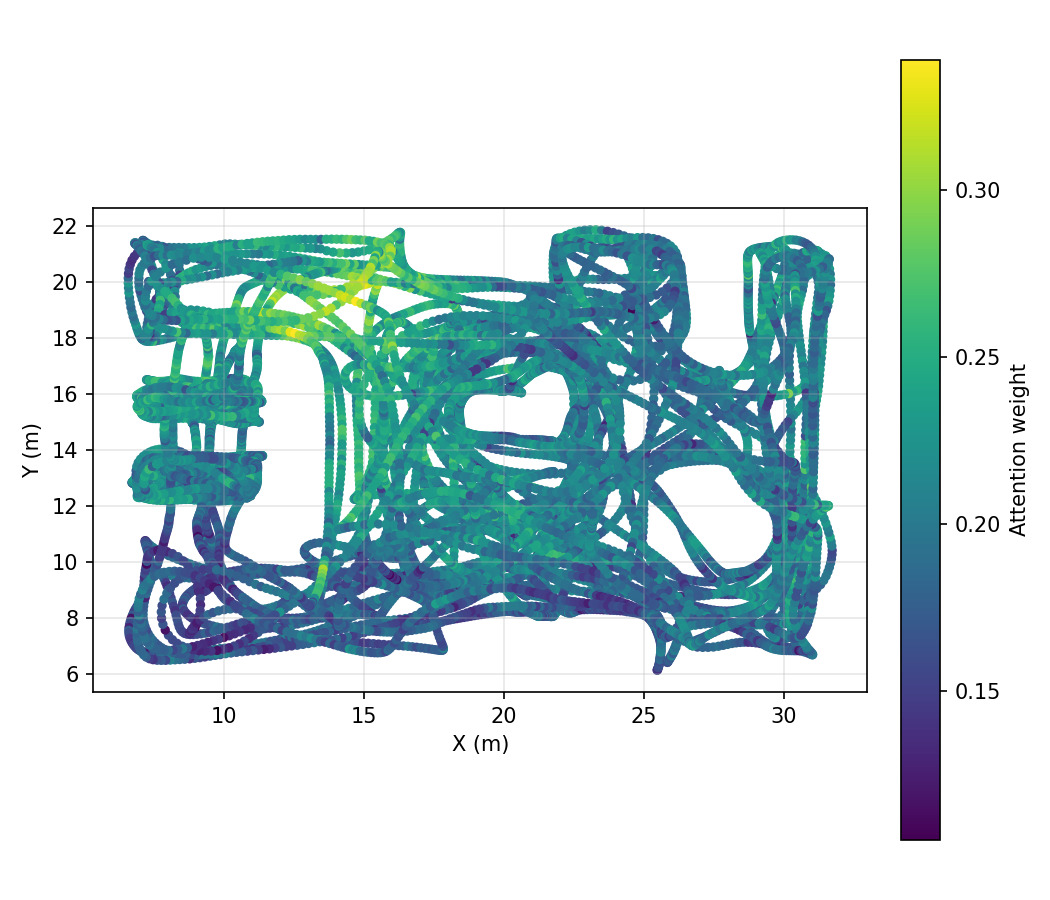}%
			\label{fig:noise_attention_weight_0}%
		}%
		\subfloat[$\bar{\boldsymbol{\alpha}}^{(0)}=0.1784$]{%
			\includegraphics[width=0.32\linewidth]{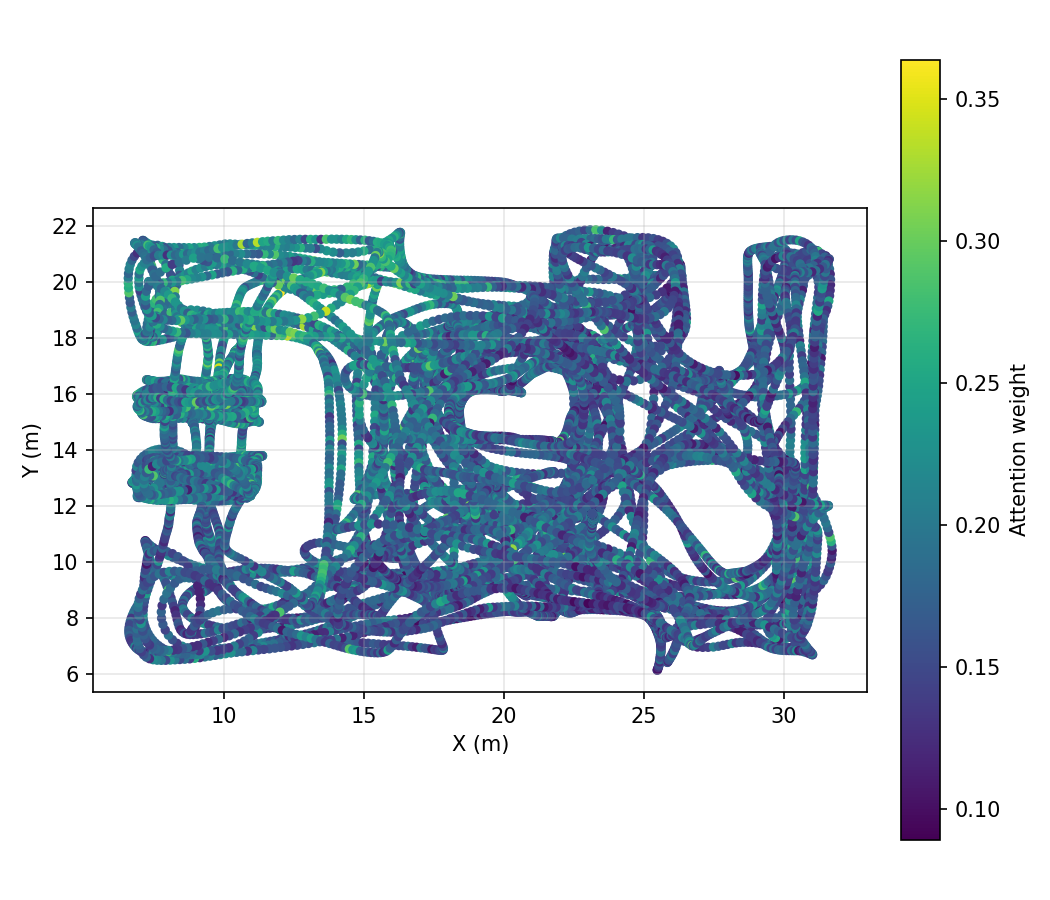}%
			\label{fig:noise_attention_weight_-10}%
		}
		\caption{Average attention weight in \gls{5G} dataset and \gls{AP} 0 for (a) clean \gls{CIR}, (b) \gls{SNR} = 0 dB, and (c) \gls{SNR} = - 10 dB.}
		\label{fig:noise_attention_weight}
	\end{figure}

	Fig. \ref{fig:noise_attention_weight} shows the average attention weight with single noisy \gls{AP} 0 for clean \gls{CIR}, \gls{SNR} = 0 dB, and \gls{SNR} = -10 dB. This figure shows how the attention layer reduces the impact of the distorted \gls{AP} by reducing its attention weight. In particular, it can be seen that the attention weight for \gls{AP} 0 has reduced from 0.2267 to 0.2057 when the \gls{SNR} degrades to 0 dB and further reduces to 0.1784 when the \gls{SNR} degrades to -10 dB. By reducing the attention weight of the distorted \gls{AP}, the output predicted location becomes more accurate as other intact \glspl{AP} take over, though some degradation occurs. In particular, the attention weight of the distorted \gls{AP} does not go to zero even with extreme noise due to the presence of clean \gls{TDoA} and \gls{ToA}-offset features. Table \ref{tab:model_comparison_resize} shows the position of the proposed LocAttMamba across all evaluated metrics.
		\begin{table*}[h]
		\centering
		\caption{Summary of the various Performance Criteria.}
		\label{tab:model_comparison_resize}
		\resizebox{\textwidth}{!}{%
			\begin{tabular}{llccc}
				\hline\hline\noalign{\smallskip}
				\textbf{Category} & \textbf{Criterion} & \textbf{5G Best} & \textbf{UWB Best} & \textbf{LocAttMamba Position} \\
				\noalign{\smallskip}\hline\noalign{\smallskip}
				
				\multirow{3}{*}{\textbf{Accuracy}} 
				& Mean error & \textbf{LocAttMamba} (1.00 m) & \textbf{LocAttMamba} (0.60 m) & Best on both \\
				& Median error & Sec5GLoc (0.83 m) & L-SwiGLU-T (0.49 m) & 2nd on both \\
				& Tail error (P90) & \textbf{LocAttMamba} (1.84 m) & \textbf{LocAttMamba} (1.03 m) & Best on both \\
				
				\noalign{\smallskip}\hline\noalign{\smallskip}
				
				\multirow{3}{*}{\textbf{Robustness}} 
				& Multiple \glspl{AP} Distorted (SNR = 5 dB) & \textbf{LocAttMamba} (5/6 cols) & \textbf{LocAttMamba} & Best except 5G/5-\gls{AP} \\
				& Single \gls{AP} Distorted (Moderate Noise) & \textbf{LocAttMamba}& \textbf{LocAttMamba} & Best on both \\
				& Single \gls{AP} Distorted (Extreme Noise) & \textbf{LocAttMamba} & L-SwiGLU-T & Best on \gls{5G}, 3rd on \gls{UWB} \\
				
				\noalign{\smallskip}\hline\noalign{\smallskip}
				
				\multirow{2}{*}{\textbf{Complexity}} 
				& Multiply-Acc. / sample & \textbf{LocAttMamba} (0.45M) & \textbf{LocAttMamba} (0.95M) & Best on both \\
				& Inference latency & Bi-LSTM & Bi-LSTM & 3rd; 4.4--16$\times$ faster than transformers \\
				\noalign{\smallskip}\hline\hline
			\end{tabular}%
		}
	\end{table*}
	\section{CONCLUSIONS}
	\label{sec:Conclusions}
	In this paper, we developed LocAttMamba, a low-complexity localization framework that processes per-\gls{AP} \gls{CIR} and time-based features with bidirectional Mamba encoders and fuses them across \glspl{AP} through a learned attention mechanism, jointly predicting the user location and its associated uncertainty. We conducted extensive experiments on two real-world \gls{5G} and \gls{UWB} datasets, where the effectiveness of the proposed design was shown. In particular, the proposed LocAttMamba achieved $1.004$ m mean error in \gls{5G} and $0.599$ m mean error in \gls{UWB}, outperforming the second-best benchmarks by $9.79\%$ and $5.82\%$, respectively, while requiring the fewest multiply-accumulate operations among all evaluated models with third-lowest inference time. Furthermore, under noisy \gls{CIR} conditions, LocAttMamba maintained its superiority in nearly all multiple-\gls{AP} and single-\gls{AP} degradation scenarios, where the attention layer was shown to adaptively down-weight distorted \glspl{AP}. Overall, the proposed LocAttMamba obtains the best or near-best performance in accuracy, robustness, and complexity simultaneously, making it well-suited to real-world deployment. \par
	As future work, graph-based fusion techniques can be considered as the spatial relationships among \glspl{AP} naturally admit a graph representation.

	\section*{Acknowledgments}
	This work has been supported by Canada Research Chairs program and NSERC Discovery grant.

\bibliographystyle{IEEEtran}
\bibliography{references}
\vfill\pagebreak
\end{document}

%% file: glossaries.tex
\newacronym{LLM}{LLM}{large language model}
\newacronym{5G}{5G}{fifth generation}
\newacronym{6G}{6G}{sixth generation}
\newacronym{XR}{XR}{extended reality}
\newacronym{V2X}{V2X}{vehicle-to-everything}
\newacronym{3-D}{3-D}{three-dimensional}
\newacronym{2-D}{2-D}{two-dimensional}
\newacronym{1-D}{1-D}{one-dimensional}
\newacronym{SSM}{SSM}{state-space model}
\newacronym{M-SSM}{M-SSM}{Mamba-based SSM}
\newacronym{S4}{S4}{structured SSM}
\newacronym{BS}{BS}{base station}
\newacronym{RSS}{RSS}{received signal strength}

\newacronym{CoT}{CoT}{chain-of-thought}
\newacronym{ToT}{ToT}{tree-of-thought}
\newacronym{GoT}{GoT}{graph-of-thought}
\newacronym{ICL}{ICL}{in-context learning}
\newacronym{GRU}{GRU}{gated recurrent unit}
\newacronym{ACNN}{ACNN}{autoencoder-based CNN}
\newacronym{KNN}{KNN}{k-nearest neighbor}
\newacronym{RAN}{RAN}{radio access network}
\newacronym{KPI}{KPI}{key performance indicator}
\newacronym{CIR}{CIR}{channel impulse response}
\newacronym{TDoA}{TDoA}{time-difference of arrival}
\newacronym{ToA}{ToA}{time of arrival}
\newacronym{MAE}{MAE}{mean absolute error}
\newacronym{M2M}{M2M}{machine-to-machine}
\newacronym{PE}{PE}{permutation entropy}
\newacronym{RMSE}{RMSE}{root mean square error}
\newacronym{MSE}{MSE}{mean square error}
\newacronym{WMA}{WMA}{weighted moving average}
\newacronym{LTI}{LTI}{linear time-invariant}
\newacronym{ZOH}{ZOH}{zero-order hold}
\newacronym{SMA}{SMA}{simple moving average}
\newacronym{LSTM}{LSTM}{long short-term memory}
\newacronym{RNN}{RNN}{recurrent neural network}
\newacronym{CNN}{CNN}{convolutional neural network}
\newacronym{ARIMA}{ARIMA}{auto regressive integrated moving average}
\newacronym{AGCRN}{AGCRN}{adaptive graph convolutional recurrent network}
\newacronym{StemGNN}{StemGNN}{spectral temporal graph neural network}
\newacronym{SDGNet}{SDGNet}{spatiotemporal dynamic graph network}
\newacronym{API}{API}{application programming interface}
\newacronym{GPU}{GPU}{graphics processing unit}
\newacronym{CSI}{CSI}{channel state information}
\newacronym{TFT}{TFT}{temporal fusion transformer}
\newacronym{STDT-Net}{STDT-Net}{spatial-temporal downsampling transformer neural network}
\newacronym{DGC}{DGC}{dynamic graph convolution}
\newacronym{UAV}{UAV}{unmanned aerial vehicle}
\newacronym{DL}{DL}{deep learning}
\newacronym{SNR}{SNR}{signal to noise ratio}
\newacronym{NLoS}{NLoS}{non-line-of-sight}
\newacronym{MLP}{MLP}{multi-layer perceptron}
\newacronym{AoA}{AoA}{angle of arrival}
\newacronym{ML}{ML}{machine learning}
\newacronym{GLU}{GLU}{gated linear unit}
\newacronym{NLP}{NLP}{natural language processing}
\newacronym{NLL}{NLL}{negative log likelihood}
\newacronym{SVM}{SVM}{Support vector machine}
\newacronym{QoS}{QoS}{quality of service}
\newacronym{IoT}{IoT}{internet of things}
\newacronym{GNSS}{GNSS}{global navigation satellite system}
\newacronym{MAC}{MAC}{multiply-accumulate}
\newacronym{P75}{P75}{75th percentile}
\newacronym{P90}{P90}{90th percentile}
\newacronym{CDF}{CDF}{cumulative distribution function}
\newacronym{UWB}{UWB}{ultra-wideband}
\newacronym{QoE}{QoE}{quality of experience}
\newacronym{mMIMO}{mMIMO}{massive multiple-input multiple-output}
\newacronym{OFDM}{OFDM}{orthogonal frequency-division multiplexing}
\newacronym{AI}{AI}{artificial intelligence}
\newacronym{LocAttMamba}{LocAttMamba}{Uncertainty-aware Mamba-Attention enabled Localization}
\newacronym{AP}{AP}{access point}
\newacronym{UV}{UV}{unmanned vehicle}
\newacronym{RSRP}{RSRP}{reference signal received power}